# Flow-based Detection of Botnets through Bio-inspired Optimisation of Machine Learning

Biju Issac, *Senior Member, IEEE,* Kyle Fryer, *Seibu Mary Jacob, Member, IEEE*
Department of Computer and Information Sciences, Northumbria University, UK
*School of Computing, Engineering and Digital Technologies, Teesside University, UK
Corresponding author: Biju Issac (e-mail: bissac@ieee.org)

*Abstract*—**Botnets could autonomously infect, propagate, communicate and coordinate with other members in the botnet, enabling cybercriminals to exploit the cumulative computing and bandwidth of its bots to facilitate cybercrime. Traditional detection methods are becoming increasingly unsuitable against various network-based detection evasion methods. These techniques ultimately render signature-based 'fingerprinting' detection infeasible and thus this research explores the application of network flow-based behavioural modelling to facilitate the binary classification of bot network activity, whereby the detection is independent of underlying communications architectures, ports, protocols and payload-based detection evasion mechanisms. A comparative evaluation of various machine learning classification methods is conducted, to precisely determine the average accuracy of each classifier on bot datasets like CTU-13, ISOT 2010 and ISCX 2014. Additionally, hyperparameter tuning using Genetic Algorithm (GA), aiming to efficiently converge to the fittest hyperparameter set for each dataset was done. The bioinspired optimisation of Random Forest (RF) with GA achieved an average accuracy of 99.85% when it was tested against the three datasets. The model was then developed into a software product. The YouTube link of the project and demo of the software developed: https://youtu.be/gNQjC91VtOI**

## I. Introduction

As the Internet grows, so does the platform on which sophisticated botnets can thrive to facilitate cybercrimes, increasingly for the purposes of financial gain. A machine can be infected with a 'bot binary' executable via traditional attack techniques, such as viruses and worms distributed through user downloads and email links, which runs silently in the background of a user machine. Once the bot binary is running, the machine becomes a 'zombie' whereby it waits for commands from a Command and Control (C&C) server, controlled by the 'botmaster', or from another bot.

Over 1400 large-scale botnet attacks targeted critical infrastructure, peaking in August-September, with Mirai, XorDDoS, Gafgyt, and hailBot dominating. China and the US were primary targets, and IoT devices, especially routers, were most vulnerable due to weak security. Mirai family controlled the largest number of devices, with QakBot and Mirai leading in Command and Control (C&C) servers, primarily in the US. UDP Flood was the prevalent attack vector, while Linux/IoT botnets, leveraging Go language, showed increased activity. Windows botnets focused on data theft and malware distribution. Predictions for 2024 include intensified infrastructure attacks, more sophisticated botnet activities, and enhanced malware concealment (Nsfocus, 2023).

A zombie machine will not generate a large amount of network traffic or processor activity until commanded to act. The traffic generated by a zombie on a network before the attack execution is for C&C server or another bot (Vacca, 2010), therefore, the proportion of the traffic generated and received by the zombie remains minimal and unnoticed.

A group of zombies under the control of a botmaster, forms a botnet to perform distributed computing tasks having control of all of the resources provided by the zombies in terms of raw computing power and network bandwidth. Botnets can be utilised for a large range of attacks, from performing Distributed Denial of Service (DDoS) attacks, to sending spam and stealing personal information to facilitate cyber crimes, such as financial fraud.

The Qrator Q1 2024 DDoS Attacks Statistics and Overview report reveals a significant growth in the largest recorded botnet, which expanded to 51,400 devices, up from 16,000 in the previous quarter. However, this number remains lower than the same period in 2023, when a botnet comprised 131,628 devices (Qrator, 2024).

Botnet usage can extend beyond DDoS attacks. They are a dynamic and heterogeneous cyber threat, even extending into modern information warfare and misinformation. Networks that control critical infrastructures, such as power and health services, are at serious risk. If a nation's infrastructure is sufficiently affected by botnet attacks, it may prompt a military response against the suspected nation. Such a response may get out of control with available nuclear weapons, therefore contributing to the Doomsday Clock being calculated at being 'still 2 minutes to midnight' (Mecklin, 2020).

As the usage of botnets as means of generating profit is increasing, especially in Cybercrime-as-a-Service (CaaS) wherein subsets of a botnet are rented out (Putman et al., 2018), botmasters must ensure that their bot evades the Intrusion Detection Systems (IDSs), to ensure that they can maintain operational persistence, allowing them to gain the trust of buyers. As a result, they can increase the potential for botnet growth to attract higher profits from buyers requiring a large amount of cumulative bandwidth and processing power.

This research aims to create a network-based IDS suitable for the behavioral differentiation between normal and malicious traffic, eliminating the need of monitoring any communication protocols, ports, architectures, and payload-based evasion mechanisms for botnet detection. Different ma-chine learning (ML) classification algorithms were evaluated using publicly available botnet datasets, with the bio-optimised tuning of classifiers' hyperparameters to ensure optimal performance. The bioinspired and optimized behavioral models were then utilized in the design of an IDS for transformation into a software product.

The rest of the paper is as follows: Section II looks into the related works followed by Section III describing the proposed work. Section IV explains the three datasets in more detail. Section V explains the feature extraction process. Section VI provides details on the important feature to be used. Section VII provides information about the program structure and the techniques. Section VIII represents the ML models, their parameters and their experiment results. Section IX displays the optimization algorithm with its search space. Section X shows the product design followed by section XI displaying the implementation for the product/software. Section XII explains the detection testing followed by Section XIII presenting the results. Section XIV informs about the future work to be carried out followed by Section XV which is the conclusion.

## II. RELATED WORK

### A. Botnets

*1) Architecture:* A bot needs to communicate with a bot-master to successfully form a botnet following centralised or decentralised topologies (Gardiner, 2014). In a centralised architecture, a classical star overlay topology is implemented, facilitating the direct communication to and from one central C&C server, generally from using the IRC protocol for a simple deployment, to Hypertext Transfer Protocol (HTTP) and HTTPS (secure HTTP) for an effective firewall traversal (Vacca, 2010). Since HTTP is the most common web protocol, running on port 80 (or 443 for HTTPS), the C&C traffic can be disguised within normal network activity.

For the architecture to be more redundant, a decentralised approach was adopted for C&C communications, whereby no single central C&C server exists, thus removing the single point of failure (Holz, 2005). A simple implementation of a decentralised architecture is by the utilisation of a distributed set of C&C servers, geographically separated to control their subset of bots, enabling load balancing and increasing the scalability potential of the botnet. If any single C&C server stops responding for any reason, the bots can redirect their traffic to other C&C servers in the botnet (Spamhaus, 2019). Additionally, portions of the botnet can then be rented or sold to buyers looking to target attacks at certain entities, increasing the profitability for bot masters by providing their botnet-as-a-service.

Peer-to-peer (P2P) bot-architecture has emerged, as a means of increasing resilience to C&C server take-down attempts. In P2P, any bot can be a C&C server, with the ability to propagate new commands to peer bots (Wang et al., 2009).

*2) Operation:* Botnets exhibit a certain operational behaviour, similar to that of a biological life cycle, in that they are born, communicate and reproduce to grow in size. To increase the size of the botnet, a host must be exploited and subsequently infected, via one or more of the active or passive propagation methods that are employed by bots (Acarali et al., 2008), such as autonomous vulnerability scanning and exploitation on a LAN (also known as a 'worm') (Gebhart, 2004), and via socially engineered spam, containing a malicious download (Gardiner et al, 2014). If the bot binary is executed on a computer, it can then use its rallying mechanisms and become a new participant in further propagation efforts of the botnet.

*3) Bot Detection Evasion Methods:* As detection methods have evolved to more effectively detect bot activity, bot binary developers have naturally responded with the implementation of various evasion methods to reduce detection and thus more effectively allow bots to maintain undetected operational persistence. Some of these methods include communication payload encryption and packet fragmentation. In a polymorphic blending attack, wherein a 'normal' profile of the traffic on the network can be established by a bot prior to communication and the characteristics of the payload can subsequently mimic the normal profile when the payload is constructed. This process operates via padding the payload with specific characters as to more closely match the normal payload byte frequency profile, synthesised via a network activity 'learning phase' of the bot, allowing the malicious traffic to blend in and thus more effectively evade specific anomaly-based detection, such as the aforementioned 'PAYL' method ((Folga, 2006; Song, 2007).

### B. Network-based Detection

Botnet detection techniques are classified into four categories: signature-based, anomaly-based, DNS-based, and data mining-based. Signature-based methods rely on known patterns, while anomaly-based approaches identify unusual network traffic but risk false positives. DNS-based techniques analyze traffic flow for irregularities in DNS queries. Data mining methods extract features from network traffic, leveraging machine learning models like Naive Bayes, Random Forest, and Support Vector Machines to classify botnet activity effectively. Supervised learning models and deep learning (e.g., HMM scores, CTU-13 dataset) show strong results in botnet detection. Unsupervised clustering methods like k-means reveal dissimilarities in traffic. IoT and smartphone-specific botnets are detected using balanced datasets and advanced machine learning techniques like SMOTE, ensemble learning, and neural networks (Mannikar et al., 2024). We will discuss some selected approaches below.

*1) Signature-based Detection:* For communication with a C&C server or other peers, a bot must be a participant in the network. The Network Intrusion Detection Systems (NIDS), such as Snort, can be utilised to help detect any such communication. The signature-based detection has been the most widely used method in practice. It works by matching the pre-defined signatures to each packet's content data; generating an alert if a match is found, deeming it as a malicious behaviour (Baker et al., 2004). However, this requires analysis of every network packet, which is, naturally, computationally intensive (Roesch, 1999) and shown in research to suffer from a large proportion of packet drops when saturated at higher network speeds (Shah

Issac, 2018). Furthermore, experimentation shows that Snort's false positive rate (FPR) can be high, with the default rule sets, rendering it difficult to analyse or trust the alerts (Shah Issac, 2018). The FPR can be calculated with the following formula in equation-1, where FP is false positive and TN is true negative:

$$\text{FPR} = \frac{\text{FP}}{\text{FP} + \text{TN}} \qquad (1)$$

*2) Anomaly-Based Detection:* Anomaly-based detection techniques, such as a payload-based IDS also known as 'PAYL' method, of n-gram analysis and modelling of the byte frequency distribution of malicious polymorphic payloads that was developed by Wang and Stolfo (2004), can be effective in detecting byte-based payload instances that exhibit a significant Mahalanobis distance from the normal distribution profile of the network traffic. The greater this distance, the more likely that the payload is anomalous and malicious due to it exceeding a dynamic threshold; as such, an alert will then be generated.

## C. Supervised Machine Learning

Due to the inadequate performance of signature-based intrusion detection systems, especially in cases where detection evasion methods are employed, other detection methods must be explored in order to ascertain techniques and methodologies that can be utilised to create a suitable IDS. Machine Learning algorithms are becoming increasingly utilised within many areas of academic research due to their ability to make classification predictions from trained models, which are generally trained in a supervised manner using labelled datasets. The classification algorithms operate via attempting to 'fit' the model to the data provided.

Chen et al. utilised conversational features from the CTU-13 botnet dataset scenarios, allowing the training of models via a range of classifiers, including DecisionTump, BayesNet and Random Forest Classifier methods (Chen et al, 2017). A feature selection algorithm, via the Random Forest Classifier's ability to rank feature importance values during training on the dataset, is used in order to help reduce the size (the dimensions) of the training dataset. Overall, the Random Forest Classifier outperformed the other classification methods under test in the evaluation, achieving a detection rate (DR) of 93.6% and false positive rate (FPR) of 0.3%. Although the FPR is favourable when compared to those present in the performance of Snort, the results are difficult to extract real world performance from due to the Detection Rate metric, which only takes into account the true positives (TP) and false negatives (FN) in the results. This may not give an accurate representation of the performance of the detection method in the context of the CTU-13 dataset scenarios, which are shown to contain a large class imbalance, whereby the malicious data accounts for an insignificant portion of the dataset. Thus the true negatives (TN) and false positives (FN) are not taken into account.

Another evaluation of classification methods for the detection of HTTP-based botnets is conducted by Dollah et al. (2018) using a research test bed with HTTP bot samples. They collected botnet traffic and merge the malicious traffic data with legitimate web browsing traffic in order to create a labelled dataset to facilitate their evaluation efforts (Dollah, 2018). Decision Tree, k-Nearest Neighbour, Naïve Bayes and Random Forest classifiers are utilised in the supervised learning mode and evaluated against Accuracy, Precision, Recall and False Positive Rate metrics. Overall, the k-Nearest Neighbours classifier resulted in the best detection performance, with an average accuracy of 92.93%. Surprisingly, this experiment showed the Decision Tree Classifier outperforming the Random Forest Classifier for each of the HTTP datasets. This is unexpected as the Random Forest Classifier is an ensemble method, whereby it utilises multiple decisions trees in order to synthesise better models. Additionally, despite the advantages of the use of multiple bot samples in the evaluation, the use of test bed-generated dataset scenarios rendered the results even more questionable. The performance cannot be easily validated by other researchers, where a publicly available dataset can be used in multiple evaluations to help compare the results from different methodologies. Moreover, the class balances or counts of the datasets are not discussed or listed, resulting in a more difficult analysis of the research as a whole, as the results cannot be correlated back to the datasets they were trained on when attempting to identify anomalies. The hyperparameters utilised for the classifiers are also not discussed or listed.

AI-based botnet detection has advanced significantly, leveraging machine learning (ML) and deep learning (DL) techniques to identify and mitigate botnet threats. Researchers have developed models such as convolutional neural networks (CNNs) and recurrent neural networks (RNNs) to analyze network traffic and detect anomalies indicative of botnet activity. A comprehensive review by Mahi (2023) examines the application of ML and DL in detecting botnet attacks on Internet of Things (IoT) devices, highlighting the evolution of detection techniques in response to sophisticated botnet strategies. Additionally, studies have focused on the resilience of detection frameworks against adversarial attacks, emphasizing the importance of robust AI models in cybersecurity (Almashor et al., 2024). The integration of AI in botnet detection continues to evolve, addressing emerging challenges in the cybersecurity landscape.

## III. PROPOSED WORK

Botnets and their network behaviour was analysed to ascertain a basis to formulate the solution. The formulation of the anomaly-based detection methods involved an investigation into the current techniques to detect malicious network flows, namely the signature-based detection methods in current NIDSs. The paper looked into the need for further techniques to be explored, i.e., to the application of ML and Neural Network (NN) in the detection of botnets in terms of their effectiveness, integrated with a strong optimisation technique.

The ML implementation was facilitated through the use of higher-level Python3 libraries via supervised learning techniques, such as Scikit-Learn (sklearn) and TensorFlow, to train, evaluate and optimise models on botnet datasets.

The feature selection/reduction was applied to identify the features in the datasets that have the largest influence in the classification of the label. Data processing algorithms can be created, if necessary, to create NetFlow datasets from normal packet capture files ('*.pcap'). Netflow is a protocol that will allow to collect the IP addresses from the network traffic. An experimental comparison of classification algorithms will be carried out to identify which of them has the most desirable performance metrics.

As labelled botnet data were used from the datasets, supervised learning can be used to train models. An algorithm is used to fit the model to some training data so that the model can predict the labels for new or unseen data. The dataset can be split into training and testing sets to allow the performance of each algorithm to be assessed. Additionally, cross-validation, such as K-Fold or ShuffleSplit (Scikit-learn, 2019a), was used to reduce any model over-fitting. A separate final testing set of untrained data can then be used to evaluate their performance.

Analysis of the performance of various supervised learning classification algorithms was done to ascertain which performed best in terms of the model's evaluation metrics, such as the F1-score, recall, precision and accuracy. A range of algorithms were used via sklearn such as Decision Trees, Random Forest, K-Neighbours, Naive Bayes, Support Vector Machine and ANN.

Figure-1 shows the project overview in terms of creating the model and processing the data.

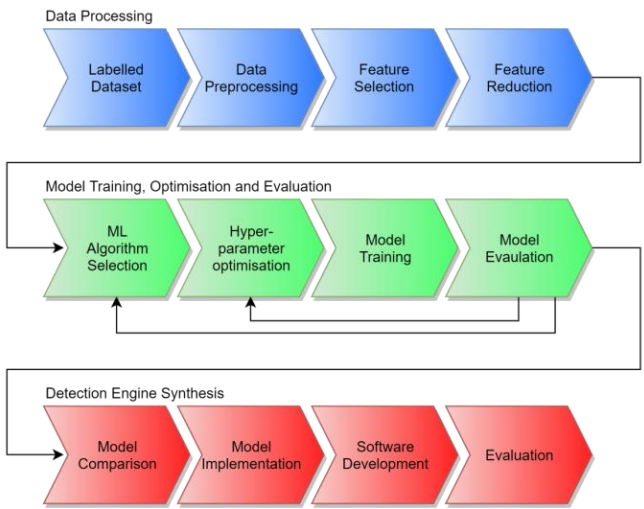

Fig. 1. The project overview

Subsequent experiments were conducted via the utilisation of multiple hidden layers in Artificial Neural network (ANN) algorithms. This is to explore the initial effects of deep learning. These algorithms are based on the theory of neurons arranged in a network in layers, whereby neurons can take multiple inputs and compute a certain output after being processed through hidden layers. This will be conducted through the use of Tensorflow (TensorFlow, 2019).

Further optimisation was carried out in the experimentation phase through tuning a classifier's hyper-parameters. After testing several combinations of hyper-parameters for each algorithm, the optimal parameters will produce the best model score. However, brute-forcing the optimisation requires a large amount of human intervention to converge on the best hyperparameter values. Algorithms such as the Exhaustive Grid Search and Randomised Parameter Optimisation exist to reduce the workload (Scikit-learn, 2019b). These methods can be compared to various biologically-inspired algorithms to evaluate the best hyper-parameter optimisation methods available for model training in this scenario. Bio-inspired algorithm can include the Genetic Algorithm (GA), Firefly Algorithm (FA), Particle Swarm Optimisation Algorithm (PSO) etc. (Gód´nez-Bautista, 2018), though we will focus on GA because of its support in existing literature.

The final model was implemented into a detection engine and a software prototype was created. Testing of the program was carried out to validate the performance of the anomaly detection engine compared to conventional NIDS methods, potentially on synthetic traffic from the datasets. Testing also included evaluating the software in terms of its quality and suitability for the desired purpose.

## IV. Datasets

Supervised ML can only be conducted when classifiers are provided with a sufficient amount of labelled data, especially in the case of low volume traffic bot communications (Khattak et al., 2014). This will allow the classification algorithms to create an accurate model, from which future predictions can be made on new unseen data. Additionally, the data format of the datasets are to be consistent to maintain the same feature headers and structure. If the data format is not correct, it can be corrected through the pre-processing stage, where manual feature extraction can be done.

### A. The Case for Network Flows

If normal traffic packet capture datasets were to remove the application layer (payload) data, signature-based detection methods are then rendered unsuitable, as the content-matching rules have no content to operate on. Although, this would address the privacy concerns but would only leave the packet headers for analysis. Using such a method, without any application data to compare known signatures, makes it very difficult to identify a bot's network communication. Hence, the flow-based features are becoming increasingly popular in traffic analysis research, as they are inherently exempt from privacy concerns by not including raw payload data, but instead extract various features from each network flow (Liu et al., 2010). Moreover, research by Haddadi, et.al shows that flow-based detection systems outperform packet-orientated rule-based detection systems (Haddadi et al., 2016). Packets belonging to the same communication connection between a source and destination are aggregated together with this method, which is then processed into a representation of the whole connection. This takes the form of a 'network flow'. Therefore, this can provide a method of representing bot communication behaviours despite the bot's employing evasion techniques like packet fragmentation and encryption.

The process of obtaining flow data is generally performed through the aggregation of raw packets into flows, then extracting uni or bi-directional features from the flow data, which is exported to a collection client for analysis. For example, this can be conducted using a flow-exporting enabled network device, such as a Cisco router exporting net flows, which are then collected by a network administrator's computer (the flow collector).

The resulting flow output can take the form of a comma separated Value (.csv) text file, which is a common file format and is supported in most data analysis software and ML libraries. Additionally, as no raw payload data is stored, such as in a .pcap file, the file sizes of network captures, in a flow format are considerably smaller and take up a fraction of the storage space. This also can help to decrease ML training time due to the smaller dataset size (Gardiner et al., 2014).

### B. Selected Datasets

Considering the factors mentioned in the previous subsection, the datasets that are most suitable are used in this work as they are collected within a physical environment (i.e. not generated through a simulation tool). They are realistic as they contain real bot traffic and are used by other researchers. Also note that multiple datasets are fundamental to sufficient evaluation of a detection system (Sommer Paxson, 2010). The following selected datasets contain a range of bot samples exhibiting their network-based communication and operational characteristics. Moreover, the resulting class balances are more representative of real scenarios, wherein the proportion of bot traffic is negligible.

Once all datasets have been pre-processed and labelled, the resulting class distributions can then be counted and analysed. This is to ensure that the datasets are consistent with the requirements to ensure that the data is representative of real world scenarios, with a realistic class balance (depending upon included traffic) and will aid in future analysis of evaluation and experimentation results that utilise these datasets. The label classes are in the form of binary categorical strings ('Normal' and 'Botnet'), which are suitable for binary classification.

*1) CTU-13:* The CTU-13 dataset was captured in 2011 at the Czech Technical University (CTU) and includes 13 scenarios, in both .pcap (botnet capture only) and labelled bi-directional network flow formats (including all background, normal and botnet traffic) (Garcia et al., 2014). It includes 7 different bot samples each exhibiting unique operational behaviour. The scenarios cover botnet usage areas including DDoS, spam, click-fraud (CF) and port scanning (PS), each of which utilises different communication protocols and architectures, such as IRC, HTTP and P2P. The total size of the 13 extended feature set network flow datasets equals 4.58GB. Table-I shows the class distribution of CTU dataset.

*2) ISOT 2010:* The ISOT 2010 dataset is a merged product of malicious and normal traffic datasets (Saad et al., 2011). The malicious traffic, from the French chapter of the Honeynet Project, comprises of activity from the Storm, Waledac and Zeus botnets. The normal background traffic datasets are the product of merged datasets from Ericsson Research and the

#### TABLE I
DATASET CLASS DISTRIBUTIONS - CTU-13

| CTU-13 Scenarios | Bots(s) | Normal Flows | Botnet Flows | Total Flows | Botnet % |
|---|---|---|---|---|---|
| 1 | Neris | 2783675 | 40961 | 2824636 | 1.45 |
| 2 | Neris | 1787181 | 20941 | 1808122 | 1.16 |
| 3 | Rbot | 4683816 | 26822 | 4710638 | 0.57 |
| 4 | Rbot | 1118496 | 2580 | 1121076 | 0.23 |
| 5 | Virut | 128931 | 901 | 129832 | 0.69 |
| 6 | Menti | 554289 | 4630 | 558919 | 0.83 |
| 7 | Sogou | 114014 | 63 | 114077 | 0.06 |
| 8 | Murlo | 2948103 | 6127 | 2954230 | 0.21 |
| 9 | Neris | 1902521 | 184987 | 2087508 | 8.86 |
| 10 | Rbot | 1203439 | 106352 | 1309791 | 8.12 |
| 11 | Rbot | 99087 | 8164 | 107251 | 7.16 |
| 12 | NSIS.ay | 323303 | 2168 | 325471 | 0.67 |
| 13 | Virut | 1885146 | 40003 | 1925149 | 2.08 |

Lawrence Berkeley National Laboratory. This data includes traffic from HTTP web browsing, gaming and P2P clients. The published dataset takes the form of a single .pcap file, 10.6GB in size. Table-II shows the class distribution of ISOT dataset.

#### TABLE II
DATASET CLASS DISTRIBUTIONS - ISOT 2010

| Bots(s) | Normal Flows | Botnet Flows | Total Flows | Botnet % |
|---|---|---|---|---|
| Zeus, Storm and Waledac | 1359595 | 113909 | 1473504 | 7.73 |

*3) ISCX Botnet 2014:* Similar to ISOT 2010, the ISCX Botnet 2014 dataset utilised a similar dataset merging methodology, with the various subsets of the included datasets being replayed over a network testbed topology, which includes a range of bot samples in the training and testing sets, to create a more generalised dataset for evaluation of bot detection methods (Beigi et al., 2014). The dataset combines 15% of the ISOT 2010 dataset, a subset of the normal traffic from the ISCX 2012 IDS and four separate bot scenarios from the CTU-13 (CIC, 2014). This dataset also takes the form of a .pcap file, with the training portion totalling 4.9GB. Table-III shows the class distribution of ISCX dataset.

#### TABLE III
DATASET CLASS DISTRIBUTIONS - ISCX BOTNET 2014

| Bots(s) | Normal Flows | Botnet Flows | Total Flows | Botnet % |
|---|---|---|---|---|
| Neris, Rbot, Virut, NSIS.ay, SMTP, Spam, Zeus | 200686 | 115890 | 316576 | 36.16 |

In the majority of the CTU-13 datasets, the proportion of botnet traffic is negligible, with only three CTU-13 scenarios (9, 10 and 11) exhibiting a more significant botnet traffic class percentage. This is expected due to the nature of the bot attack behaviours that are exhibited. These specific attacks generally operate at volume in order to allow the botnet to sufficiently propagate and perform their malicious actions at a larger scale, thus producing more traffic than that of solely periodic C&C communication. Scenario 9 has attacks consisting of spamming, click-fraud and port scanning through 10 Neris bot instances, scenario 10 has UDP and ICMP DDoS attacks by Rbot using 10 bots instances and scenario 11 has 3 bot instances, respectively as in Table-I (Garcia et al., 2014). CTU-13's scenario 7 (Sogou) exhibits the lowest botnet class percentage with only 0.06%, which is coupled with the fact that this dataset contains second lowest number of total flows; as a result, only 63 flows are in the botnet class. This is indicative of the challenges of botnet detection, as bots do not communicate at volume Gu et al., 2008), which highlights the importance of the requirement that datasets contain a large volume of normal traffic and are conducted over a long period of time, to ensure that sufficient botnet communication and attack behaviour can be recorded.

The largest bot class proportion is present in the ISCX Botnet 2014 dataset, possibly due to the fact that it includes multiple bot samples. This results in cumulative bot attack behaviour flows, increasing the overall proportion of bot traffic. This dataset will still be utilised in the evaluation in order to understand the effect of models that are trained on more bot behaviours, allowing an evaluation of the effect of behavioural model generalisation when applied to normal traffic.

CTU-13 and ISOT 2010 datasets have been produced towards providing a range of realistic and representative bot behaviours, where the proportion of botnet traffic compared to normal traffic is lower, as they intend to operate without causing detectable noise on the network or impact performance when they are not conducing volumetric attacks (Khattak et al, 2014). Thus there is an inherent realistic class imbalance in the majority of the datasets, which is a desirable characteristic when they will be applied in an evaluation of behavioural modelling classification algorithms. The normal and bot flow distribution of the three datasets are shown in Figure-2.

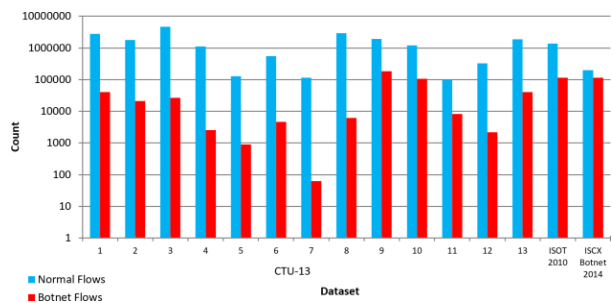

Fig. 2. The normal and bot flow across all datasets

## V. Feature Extraction

The CTU-13 is already in the required network flow format, with both base and extended feature extracted datasets publicly available. However, this is not the case for the ISOT 2010 and ISCX 2014 datasets, which are available as .pcap files only. Therefore, they must be processed via feature extraction, into network flow format, with the same feature set present in the extended CTU-13, as to maintain consistency.

### A. Feature Sets and Descriptions

Two main feature sets are present in the CTU-13, namely the base and extended network flow datasets, which include a total of 15 and 33 features respectively (including the 'Label' field) (Garcia et al., 2014). The base feature set includes both source and destination IP addresses and port numbers, along with basic total bytes and packet counts that occurred during the connection. In contrast, the extended feature set expands connection data analysis with the presence of additional feature fields. These fields intend to provide more specific information about both the source and destination total bytes, packets, rate, application layer bytes, time-to-live (TTL) values, estimated number of hops from the flow collector, along with TCP-extracted flow information such as source and destination window sizes, round-trip-time (RTT) and connection setup times. The full feature descriptions for both the base and extended datasets are available in Table-IV, and these are derived from the 'ra' man page (Bullard, 2020). The extended array of more detailed features available in the extended datasets can cover more sophisticated details of specific network activities. In a comparative evaluation by Haddadi et.al between the base and extended CTU-13 datasets (Haddadi et al., 2016), facilitated by 10-fold cross-validation of the C4.5 decision tree algorithm on each dataset, it is shown that the extended feature sets yield higher detection rate performance. They concluded that the specific categories present in the extended feature set are more suitable for bot communication detection. Therefore, the extended feature set will be utilised in this proposed work for experimentation and evaluation. This feature set is also more suitable for feature selection as compared to the base feature set.

### B. Process

As the CTU-13 dataset is already available with an extended feature set, only the ISOT 2010 and ISCX Botnet 2014 datasets are required to be specifically processed, with the feature set matching that of CTU-13. The feature extraction process, from raw packet capture files to network flow data, is conducted via the use of the Argus network flow collector and processor to create flow records (Argus, 2012). This data can then be passed into the Argus Client command 'ra' which reads argus data, where the delimiter character is set to a comma (-c ','), to present the data in .csv format with the correct feature set (Bullard, 2020). Additionally, the required fields can be defined using -s command-line option to align with those present in the CTU-13 extended network flow feature set. Bash scripts were developed to process the ISOT 2010 and ISCX Botnet 2014 packet capture datasets into the required network flow datasets.

For example, the script in Table-V can be used for extended network flow feature extraction, from a raw packet capture file, used for the ISCX Botnet 2014 dataset.

### C. Dataset Preparation and Pre-Processing

To facilitate ML in Python 3.7, using the sklearn library (Pedregosa et al., 2011), the datasets must first be loaded

TABLE IV
EXTENDED DATASET FEATURE DESCRIPTIONS [**Bullard**]

| Base-Feature | Description | Extended-Feature | Description |
|---|---|---|---|
| StartTime | The flow start time | SrcAddr | The source IPv4/IPv6 address |
| Dur | Total flow duration | DstAddr | The destination IPv4/IPv6 address |
| Proto | The protocol used | Proto | The protocol used |
| SrcAddr | The source IPv4/IPv6 address | Sport | The source port number |
| Sport | The source port number | Dport | The destination port number |
| Dir | Direction of the flow (uni-directional, bi-directional or unknown) | State | The state of the connection |
| DstAddr | The destination IPv4/IPv6 address | ✓sTos | Source ToS value |
| Dport | The destination port number | ✓dTos | Destination ToS value |
| State | The state of the connection | ✓SrcWin | Source TCP window size |
| sTos | Source ToS value | ✓DstWin | Destination TCP window size |
| dTos | Destination ToS value | ✓sHops | The number of estimated IP hops from the source to the flow collector |
| TotPkts | Total number of packets | ✓dHops | The number of estimated IP hops from the destination to the flow collector |
| TotBytes | Total number of bytes transferred | StartTime | The flow start time |
| SrcBytes | Total number of source-only bytes transferred | LastTime | The flow end time |
| Label | The label of the traffic (Normal or Botnet) | ✓sTtl | The Source to Destination TTL value |
| - | - | ✓dTtl | The Destination to Source TTL value |
| - | - | ✓TcpRtt | The TCP Round-Trip Time (SYN-ACK to ACK-DAT) |
| - | - | ✓SynAck | TCP Connection setup time; between SYN and the SYN-ACK |
| - | - | ✓AckDat | TCP Connection set up time; between SYN-ACK and ACK |
| - | - | ✓SrcPkts | Total number of source-only packets transferred |
| - | - | ✓DstPkts | Total number of destination only packets transferred |
| - | - | ✓SrcBytes | Total number of source-only bytes transferred |
| - | - | ✓DstBytes | Total number of destination only bytes transferred |
| - | - | ✓SAppBytes | Source to Destination Application bytes transferred |
| - | - | ✓DAppBytes | Destination to Source Application bytes transferred |
| - | - | ✓Dur | Total flow duration |
| - | - | ✓TotPkts | Total number of packets transferred |
| - | - | ✓TotBytes | Total number of bytes transferred |
| - | - | ✓TotAppByte | Total number of application bytes transferred |
| - | - | ✓Rate | Total number of packets per second |
| - | - | ✓SrcRate | Source packets per seconds |
| - | - | ✓DstRate | Destination packets per second |
| - | - | Label | The label of the traffic (Normal or Botnet) |

TABLE V
SAMPLE SCRIPT FOR EXTENDED NETWORK FLOW FEATURE EXTRACTION FOR ISCX BOTNET 2014 DATASET

| Line no. | Linux Command |
|---|---|
| 1 | #!/bin/bash |
| 2 | echo "Starting ISCX pcap Feature Extraction into Extended Network Flow format ..." |
| 3 | sudo argus -r ISCX Botnet-Training.pcap -w ISCX Training.argus |
| 4 | ra -r ISCX_Training.argus -c ',' -n -s saddr daddr proto sport dport state stos dtos swin dwin shops dhops stime ltime sttl dttl tcprtt synack ackdat spkts dpkts sbytes dbytes sappbytes dappbytes dur pkts bytes appbytes rate srate drate label > Netflow Extended/ISCX Training.binetflow |
| 5 | echo "Feature extraction complete!" |

from the .csv file into the program's Random-access memory (RAM), into a format suitable and supported by sklearn. This is achieved via the Pandas library with its DataFrame data structure object (Pandas, 2014a), chosen for the wide range of object attributes and data manipulation functions available. Dataset loading is conducted via pandas.read_csv (Pandas,

2014b), which returns the data as a DataFrame object.

Additionally, the CTU-13's labels contain verbose descriptions of the exact network activity, such as flow=From-Botnet-V50-4-TCP-WEB-Established-SSL and flow=From-Normal-V50-Stribrek, which are not suitable for categorical binary classification due to the range of descriptions. Therefore, during the pre-processing iterations, the 'Label' feature vector fields for CTU-13 were normalised to only contain 'Normal' and 'Botnet' categorical binary string labels. In the case of the ISOT 2010 and ISCX Botnet 2014, which are not directly labelled due to their packet capture file format, labels must be added to identify the network flow traffic, depending upon the identified infected machines' IP addresses, detailed in each dataset's documentation.

## VI. FEATURE SELECTION/REDUCTION

Through a process of feature selection and reduction, the dataset dimensionality can be reduced, thus reducing ML classifier fitting times, whilst retaining the more important features, without inducing a significant negative impact on classifier performance (Fatima et al., 2019; Khan et al., 2019). As a side effect of such a process being carried out, feature importance metrics can be computed and subsequently help to provide an avenue down and the various behavioural characteristics can be analysed. This will help to gain an understanding of how different bot behaviours can be identified via the specific subsets through network flow features, providing the most variance in the classification process.

### A. Manual Feature Exclusion

Features vectors from the dataset that are not essential were removed from the training data, such as the network flow metadata, dates, connection start and end time, along with the source and destination IP addresses. This is a cardinal part of data preprocessing for the design of the model's capabilities, as feature vectors that are manually excluded ensure that the classifiers are trained, evaluated and operated in a desirable way. The aim is to develop a detection system that is completely independent of the port numbers and protocols that are being utilised by bots in the network flow-based communications data (Mahmoud et al., 2015). Such considerations must be applied in the context of the range of botnet detection evasion methods that are being increasingly utilised and therefore must be taken into account, such as the dynamic port fluxing methods (Beigi et al., 2014). This can also allow to train a model to allow generalisation to only the behavioural feature vectors (Haddadi et al., 2016).

When these considerations were applied and the related features were excluded, the original extended dataset feature set was reduced to 24 instead of 33 (shown with a checkmark in table-IV), which only contains integer and float data types. This exclusion takes place within the dataset splitting function, via defining the feature labels that are to be maintained, as a result of the manual feature exclusion process, in an array and then utilising the 'pandas.DataFrame.loc' attribute (Pandas, 2014c), allowing the resulting 'X' feature vector DataFrame to be returned containing only the relevant features.

### B. Computing Feature Importance

Each feature in the feature set exhibits a 'weight' in the classification of data and thus weaker or irrelevant features can be removed to reduce the dimensionality of the dataset. This allows training time to significantly reduce whilst keeping the classifier performance to acceptable values (Scikit-Learn, 2019b). One of the most prominent methods is Random Forest (RF) classifier that proves to be effective and will be utilised in this work (Chen et al., 2017). Further exploration of feature selection methods with a comparative evaluation was carried out by Pektas, & Acarman (2017), whereby tree-based feature selection via a RF provides the highest F1-score of 0.99/1.0 with 10-fold cross-validation on ISOT dataset.

In this work, the feature ranking was performed using RF in sklearn, fitting the dataset results in a model object that contains a feature_importances_ attribute array. In this methodology, only the top 15 features of each dataset were retained for training and testing of each classifier's performance.

### C. Feature Importance Analysis

Once the feature selection methodology was applied to each dataset, the feature importance values were collected, averaged and presented in Figure-3. It is clear from this analysis that sTos, dTos, ackDat and DstWin result in the lowest average feature importance values.Therefore, we can conclude that the general botnet network communication behaviour in these particular weak features are close to the normal network traffic behaviour and thus cannot be used to differentiate the traffic well. On the other hand, botnet behaviour can be more sufficiently characterised by the high feature importance values present in the sTtl, SrcBytes, SrcWin and TotBytes features.

This analysis proves that the feature selection should not be done on an averaged value basis, and instead should be conducted on a per-dataset basis. This is because different bots exhibit different operational behaviours, network communications and types of attacks. When conducting the investigation, in CTU-13's scenario 2 (Spam and Port Scanning behaviours using HTTP communication), SrcWin proved to be the most important feature with a value of 0.19, whereas, in CTU-13's scenario 10 (UDP DDoS attacks with an IRC communication protocol), SrcWin is one of the weakest features with a value of 0.004.

Computing feature importance can help in the dataset and bot characteristic analysis and thus can also help remove features with minimal importance in the classification process. As a result, the training and classifier evaluation times can be reduced with only a small classifier performance deterioration.

## VII. Product Methodology

### A. Dataset Splitting

To split the datasets into an $X$ feature vector DataFrame and $y$ label vector array for training, a function is used to return both vectors for any given network flow dataset. Feature selection is incorporated in this dataset splitting functionality, returning 15 features with the highest importance socres in the $X$ feature vector for each dataset.

### B. Stratified 10-Fold Cross Validation

Instead of the normal 70:30 training-testing single fitting, which can result in an inaccurate evaluation of actual classifier performance and is especially weak when utilised in class imbalanced datasets (Pektas Acarman, 2017), this methodology utilised stratified k-fold cross-validation for an accurate classifier performance.

This method operates by splitting the dataset into k folds, whereby $k - 1$ folds are used for training and the remaining fold is used for testing; this is repeated $k$ times such that each fold is used for testing at least once and for training $k - 1$ times. Additionally, it is stratified in the splitting behaviour which means that the class balances are preserved in each fold of dataset. In this work, $k$ is set to 10 to allow a more thorough evaluation of the classifier performance on the class imbalanced datasets.

As the sklearn cross_validate function includes a multithreading argument n_jobs, each fold can be allocated with a CPU core to allow the 10 folds for a single dataset to run in parallel.

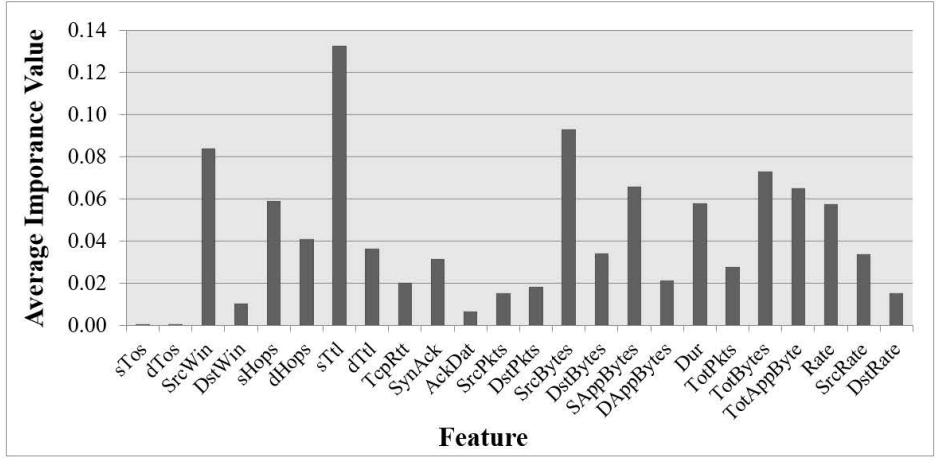

Fig. 3. Computed feature importance values; averaged across all datasets

## C. Evaluation Metrics

Due to the class imbalances present in the datasets, this must be taken into account when choosing the metrics which are to be used in the evaluation of binary classification performance of each classification method. Due to these class imbalances, accuracy alone will not be a suitable metric due to the positive class samples accounting for the minority of the total samples. The accuracy metric takes into account the True Negative (TN) results which causes the TN predictions to have a large weight on the accuracy metric. Accuracy can be calculated to compare the metrics as in equation-2:

$$\text{Accuracy} = \frac{(\text{TP} + \text{TN})}{(\text{TP} + \text{TN} + \text{FP} + \text{FN})} \quad (2)$$

Therefore, classification performance metrics are required that do not take into account the TN predictions and thus are weighted to the prediction performance on the minority class. To this end, the F1-Score is calculated as the harmonic mean between the Precision and Recall metrics, as shown below, which do not take into account the TN predictions. So it will be good to look at both accuracy and F1-score.

Precision is a function of the TP predictions divided by the total positive samples, allowing the characterisation of a classifier's ability to correctly classify the positive samples as in equation-3.

$$\text{Precision} = \frac{\text{TP}}{(\text{TP} + \text{FP})} \quad (3)$$

Recall is a function of the TP predictions divided by the sum of the TP and FN samples, allowing the characterisation of a classifier's ability to correctly classify the relevant samples as in equation-4.

$$\text{Recall} = \frac{\text{TP}}{(\text{TP} + \text{FN})} \quad (4)$$

The F1-Score values range between 0.0 and 1.0, wherein a higher value is indicative of better classification performance as in equation-5.

$$\text{F1} - \text{Score} = 2 \times \frac{(\text{Precison} \times \text{Recall})}{(\text{Precision} + \text{Recall})} \quad (5)$$

Additionally, as each classifier is using different algorithms, the average fitting time of each fold in the cross-validation can be gathered to evaluate the computational time required to fit each of the datasets. This will help to characterise how effective each algorithm is in producing models in a suitable time.

## VIII. MACHINE LEARNING TECHNIQUES

Binary classification of the data can be achieved by utilising supervised machine learning classifiers, with the labelled network flow datasets, to create models that are suitable for predictions on future unseen data. The ML classifiers were chosen due to their ability to correctly process the data and its various types, which was initially explored in the WEKA software (Frank et al., 2016) as in Table-VI and were also chosen from the literature review. The classifiers that did not support the datasets were marked as unavailable and therefore unsuitable for the type of data being used.

TABLE VI
INITIAL WEKA RESULTS

| ML Classifier | F1 score |
|---|---|
| Naïve Bayes | 0.231 |
| KNN | 0.521 |
| Ada Boost | 0.857 |
| Decision Tree | 0.986 |
| Random Forest | 0.987 |
| SVC (SMO) | 0.533 |

This section discusses the ML algorithms utilised to identify the ones with best performance. This will be fed into the optimization algorithm for further improvement. The section also gives the results based on the default value for each classifier, along with their parameters that are considered within the program.

## A. Gaussian Naïve Bayes

Based on the Bayes theorem, the Gaussian Naïve Bayes (GNB) classifier operates by assuming that the features are independent of each other and their values are numerical with normal distribution (John Langley, 1995). The training data can be split by class ($y$) to allow the analysis of the mean ($\mu_y$) and corrected standard deviation ($\sigma_y^2$) of each feature vector ($x_i$) distribution, allowing the probability density of a class to be calculated using the equation-6:

$$P(x_i|y) = \frac{1}{\sqrt{2\pi\sigma_y^2}} exp(-\frac{(x_i - \mu_y)^2}{2\sigma_y^2}) \qquad (6)$$

Table-VII shows the metrics output for GNB, with highest F1-score of 0.7725 for CTU-13-scenario 3, along with fit-time as 8.57 sec, precision as 0.6336, recall as 0.9893 and accuracy as 0.7730. When all the scenarios are considered, the performance is poor.

### TABLE VII
### GAUSSIAN NAÏVE BAYES

| Dataset | | Fit Time (s) | Precision | Recall | F1 score | Accuracy |
|---|---|---|---|---|---|---|
| CTU-13 | 1 | 6.45 | 0.5089 | 0.5991 | 0.5503 | 0.5622 |
| | 2 | 3.34 | 0.5081 | 0.6523 | 0.5712 | 0.5768 |
| | 3 | 8.57 | 0.6336 | 0.9893 | 0.7725 | 0.7730 |
| | 4 | 1.87 | 0.5020 | 0.7175 | 0.5907 | 0.5988 |
| | 5 | 0.18 | 0.5055 | 0.6896 | 0.5834 | 0.5960 |
| | 6 | 0.86 | 0.5130 | 0.8418 | 0.6375 | 0.6394 |
| | 7 | 0.16 | 0.5002 | 0.5549 | 0.5261 | 0.6055 |
| | 8 | 9.87 | 0.5021 | 0.7559 | 0.6034 | 0.6199 |
| | 9 | 5.71 | 0.5431 | 0.5486 | 0.5459 | 0.5541 |
| | 10 | 2.59 | 0.6055 | 0.8332 | 0.7013 | 0.7016 |
| | 11 | 0.15 | 0.5922 | 0.7969 | 0.6795 | 0.6799 |
| | 12 | 0.64 | 0.5033 | 0.5425 | 0.5222 | 0.5516 |
| | 13 | 6.13 | 0.5109 | 0.5698 | 0.5387 | 0.5464 |
| ISOT | | 4.23 | 0.1386 | 0.5000 | 0.2171 | 0.2185 |
| ISCX | | 0.61 | 0.6969 | 0.5779 | 0.6318 | 0.6313 |

### B. Decision Tree

In Decision Tree (DT) classifier a tree is constructed by nodes and vertices where conditional tests are synthesised at each node. This allows an effective split between the branches that represent the classes of data present, resulting in a tree-like structure of decisions. The rules for nodes/branches get more complicated with the depth of the tree (Decision Trees, 2019). Unseen feature vectors can then be fed to the root node and propagated down to the leaf nodes that denote the relevant class. The classification criteria can be denoted by the formula in equation-7, which is then utilised for calculating gini, entropy and misclassification (Decision Trees, 2019).

$$P_{mk} = \frac{1}{N_m} \sum_{x_i \in R_m} I(y_i = k) \qquad (7)$$

where $m$ is the node in the tree, $R_m$ is the region, $N_m$ are the observations and $x$ is the data-point.

Table-VIII shows the metrics output for DT, with highest F1-score of 0.9990 for CTU-13-scenario 10, along with fit-time as 10.06 sec, precision as 0.9994, recall as 0.9987 and accuracy as 0.9994.

Table-IX shows the hyperparameters to be considered to optimise the performance of DT. The default parameters used are ['gini', 'best', 2, 1, 0.0, None].

### TABLE VIII
### DECISION TREE

| Dataset | | Fit Time (s) | Precision | Recall | F1 score | Accuracy |
|---|---|---|---|---|---|---|
| CTU-13 | 1 | 16.26 | 0.9711 | 0.9688 | 0.9699 | 0.9909 |
| | 2 | 12.09 | 0.9827 | 0.9846 | 0.9836 | 0.9933 |
| | 3 | 55.63 | 0.9976 | 0.9986 | 0.9981 | 0.9987 |
| | 4 | 5.17 | 0.9769 | 0.9844 | 0.9806 | 0.9940 |
| | 5 | 0.45 | 0.9567 | 0.9566 | 0.9566 | 0.9772 |
| | 6 | 2.66 | 0.9908 | 0.9980 | 0.9944 | 0.9974 |
| | 7 | 0.40 | 0.7493 | 0.7653 | 0.7572 | 0.8716 |
| | 8 | 21.27 | 0.9639 | 0.9741 | 0.9690 | 0.9956 |
| | 9 | 14.94 | 0.9696 | 0.9690 | 0.9693 | 0.9840 |
| | 10 | 10.06 | 0.9994 | 0.9987 | 0.9990 | 0.9994 |
| | 11 | 0.53 | 0.9992 | 0.9984 | 0.9988 | 0.9994 |
| | 12 | 2.16 | 0.8952 | 0.9187 | 0.9068 | 0.9579 |
| | 13 | 13.40 | 0.9774 | 0.9809 | 0.9792 | 0.9933 |
| ISOT | | 19.50 | 0.9937 | 0.9829 | 0.9883 | 0.9949 |
| ISCX | | 2.46 | 0.9727 | 0.9781 | 0.9754 | 0.9747 |

### TABLE IX
### DECISION TREE - PARAMETER SELECTION

| Parameter | Search Space |
|---|---|
| Criterion | ['gini', 'entropy'] |
| Splitter | ['best', 'random'] |
| Min samples split | (low = 2, high = 5) |
| Min samples leaf | (low = 1, high = 4) |
| Min weight fraction leaf | (low = 0.0, high = 0.1) |
| Class weight | ['balanced', none] |

### C. Random Forest

The Random Forest (RF) classifier utilises training of multiple DT classifiers on subsets of the dataset to allow model averaging for an improved performance. By constructing multiple individual trees, the feature importance values can be more accurately calculated at each decision node, ultimately reducing the variance (Scikit-Learn, 2019c). The cumulative boosting estimation of the trees within the algorithm can be denoted by equation-8:

$$F_M(x_i) = \sum_m h_m(x_i) \qquad (8)$$

where $M$ is the iteration, $x_i$ is the input datapoint and $h_m$ are the sub-estimators.

Table-X shows the metrics output for RF, with highest F1-score of 0.9992 for CTU-13-scenario 10, with fit-time as 21.66 sec, precision as 0.9996, recall as 0.9988 and accuracy as 0.9996.

Table-XI shows the hyperparameters to be considered to optimise the performance of RF. The default parameters used are [100, 'gini', 2, 1, 0.0, None].

### D. AdaBoost

Similar to the discussion behind the RF Classifier, the AdaBoost classifier utilises a boosting algorithm to improve the performance of a base algorithm and in the sklearn implementation, this is a DT classifier. At each fitting interval, the classifier is fitted on the dataset but modifies the weights of the incorrectly classified nodes from the previous fitment to allow those particular incorrect cases to be considered more prominently (Scikit-Learn, 2019c). This is a similar method to error-rate reduction as used in Artificial Neural Networks. But

### TABLE X
### RANDOM FOREST

| Dataset | | Fit Time (s) | Precision | Recall | F1 score | Accuracy |
|---|---|---|---|---|---|---|
| CTU-13 | 1 | 34.01 | 0.9818 | 0.9698 | 0.9759 | 0.9977 |
| | 2 | 27.59 | 0.9915 | 0.9854 | 0.9884 | 0.9981 |
| | 3 | 80.07 | 0.9991 | 0.9991 | 0.9991 | 0.9996 |
| | 4 | 13.82 | 0.9863 | 0.9833 | 0.9848 | 0.9983 |
| | 5 | 0.87 | 0.9895 | 0.9496 | 0.9692 | 0.9901 |
| | 6 | 7.21 | 0.9949 | 0.9980 | 0.9964 | 0.9993 |
| | 7 | 0.68 | 0.9113 | 0.7987 | 0.8513 | 0.9799 |
| | 8 | 75.71 | 0.9725 | 0.9729 | 0.9727 | 0.9993 |
| | 9 | 20.96 | 0.9720 | 0.9800 | 0.9760 | 0.9907 |
| | 10 | 21.66 | 0.9996 | 0.9988 | 0.9992 | 0.9996 |
| | 11 | 0.77 | 0.9996 | 0.9985 | 0.9991 | 0.9997 |
| | 12 | 4.31 | 0.9641 | 0.9096 | 0.9361 | 0.9889 |
| | 13 | 33.58 | 0.9839 | 0.9836 | 0.9838 | 0.9980 |
| ISOT | | 31.11 | 0.9929 | 0.9846 | 0.9887 | 0.9953 |
| ISCX | | 4.822 | 0.9912 | 0.9930 | 0.9921 | 0.9913 |

### TABLE XI
### RANDOM FOREST - PARAMETER SELECTION

| Parameter | Search Space |
|---|---|
| N estimators | (low = 10, high = 200) |
| Criterion | ['gini', 'entropy'] |
| Min samples split | (low = 2, high = 5) |
| Min samples leaf | (low = 1, high = 4) |
| Min weight fraction leaf | (low = 0.0, high = 0.1) |
| Class weight | ['balanced', none] |

by using sequences of classifier instances to reduce the error rate in the next iteration, the number of classifier instances to be used in the training can be defined in the n_estimators hyperparameter, which defaults to 50 trees. Standard Adaboost algorithm can be denoted by equation-9 (Li et al., 2008):

$$H_{\text{Ada}}(I) = \text{sign}(\sum_{t=1}^{T} \alpha_t h_t(I)) \qquad (9)$$

where $I$ is the datapoints/labels within the dataset, $\alpha_t$ is the weight of the classifier, $h_t$ is the classifier and $T$ is the number of cycles.

Table-XII shows the metrics output for Adaboost, with highest F1-score of 0.9991 for CTU-13-scenario 11, along with fit-time as 11.22 sec, precision as 0.9996, recall as 0.9985 and accuracy as 0.9991.

### TABLE XII
### ADABOOST

| Dataset | | Fit Time (s) | Precision | Recall | F1 score | Accuracy |
|---|---|---|---|---|---|---|
| CTU-13 | 1 | 396.06 | 0.9593 | 0.9142 | 0.9362 | 0.9565 |
| | 2 | 303.21 | 0.9906 | 0.9683 | 0.9793 | 0.9889 |
| | 3 | 781.93 | 0.9969 | 0.9972 | 0.9971 | 0.9977 |
| | 4 | 141.87 | 0.9645 | 0.9071 | 0.9349 | 0.9477 |
| | 5 | 11.55 | 0.9757 | 0.9373 | 0.9561 | 0.9767 |
| | 6 | 94.31 | 0.9918 | 0.9951 | 0.9935 | 0.9964 |
| | 7 | 10.06 | 0.9415 | 0.7511 | 0.8356 | 0.9618 |
| | 8 | 748.56 | 0.9652 | 0.9707 | 0.9679 | 0.9944 |
| | 9 | 229.98 | 0.9442 | 0.9377 | 0.9409 | 0.9551 |
| | 10 | 301.55 | 0.9991 | 0.9981 | 0.9986 | 0.9990 |
| | 11 | 11.22 | 0.9996 | 0.9985 | 0.9991 | 0.9991 |
| | 12 | 37.52 | 0.9551 | 0.7254 | 0.8245 | 0.8709 |
| | 13 | 253.61 | 0.9684 | 0.9470 | 0.9576 | 0.9714 |
| ISOT | | 196.59 | 0.9693 | 0.9511 | 0.9601 | 0.9665 |
| ISCX | | 42.46 | 0.9802 | 0.9813 | 0.9808 | 0.9800 |

Table-XIII shows the hyperparameters to be considered to optimise the performance of Adaboost. The default parameters used are [50, 1.0, 'SAMME.R', None].

### TABLE XIII
### ADABOOST - PARAMETER SELECTION

| Parameter | Search Space |
|---|---|
| N estimators | (low = 5, high = 100) |
| Learning rate | (low = 0.1, high = 0.1) |
| Algorithm | ('SAMME', 'SAMME.R') |
| Random state | (low = 1, high = 50) |

### E. Linear Support Vector Machines

Support Vector Machines (SVMs) are a method of classifying data by fitting a hyperplane to the feature vectors, creating a split in the data where it best determines the data labels can be effectively split. As a result, they are useful for datasets that have large dimensionality. A linear kernel is applied to the SVM classifier, allowing faster operation due to a simpler function to fit the data (Support Vector Machines, 2019).

The equation-10 below shows the primal problem by utilising the hinge loss:

$$\max_{w,b} \frac{1}{2} w^T w + C \sum_{i=1} \max(0, y_i(w^T \phi(x_i) + b)) \qquad (10)$$

where $x$ is the datapoint, $\phi$ is the identify function and $C$ is the penalty term.

Table-XIV shows the metrics output for Linear SVC, with highest F1-score of 0.9555 for CTU-13-scenario-10; along with fit-time as 781.95 sec, precision as 0.9363, recall as 0.9756 and accuracy as 0.9559.

### TABLE XIV
### LINEAR SVC

| Dataset | | Fit Time (s) | Precision | Recall | F1 score | Accuracy |
|---|---|---|---|---|---|---|
| CTU-13 | 1 | 1715.69 | 0.5831 | 0.6131 | 0.5977 | 0.6106 |
| | 2 | 626.35 | 0.5458 | 0.6945 | 0.6113 | 0.6173 |
| | 3 | 1514.78 | 0.4971 | 0.4956 | 0.4963 | 0.4965 |
| | 4 | 931.57 | 0.5284 | 0.6992 | 0.6019 | 0.6101 |
| | 5 | 35.88 | 0.5019 | 0.5552 | 0.5286 | 0.5400 |
| | 6 | 407.23 | 0.6498 | 0.9131 | 0.7593 | 0.7615 |
| | 7 | 23.19 | 0.4997 | 0.4954 | 0.4975 | 0.5726 |
| | 8 | 886.48 | 0.5308 | 0.7854 | 0.6335 | 0.6508 |
| | 9 | 625.14 | 0.6884 | 0.7862 | 0.7341 | 0.7452 |
| | 10 | 781.95 | 0.9363 | 0.9756 | 0.9555 | 0.9559 |
| | 11 | 28.82 | 0.9277 | 0.9678 | 0.9473 | 0.9479 |
| | 12 | 107.18 | 0.5036 | 0.5192 | 0.5113 | 0.5401 |
| | 13 | 556.60 | 0.5845 | 0.6752 | 0.6266 | 0.6356 |
| ISOT | | 568.31 | 0.6283 | 0.6634 | 0.6454 | 0.6497 |
| ISCX | | 111.19 | 0.7135 | 0.6634 | 0.6876 | 0.6870 |

Table-XV shows the hyperparameters to be considered to optimise the performance of Linear SVC. The default parameters used are ['squared_hinge', 1e-4, 1.0].

TABLE XV
LINEAR SVC - PARAMETER SELECTION

| Parameter | Search Space |
|---|---|
| Loss | ['hinge', 'squared hinge'] |
| Tol (tolerance) | (low = 1e-5, high = 0.1) |
| C (Regularisation Parameter) | (low = 1, high = 5) |

*F. K Nearest Neighbors*

The k-Nearest-Neighbours (KNN) classifier is an algorithm that calculates the distance of a data sample point from the nearest k number of data sample neighbours in the dataset's trained feature space. The nearest k labels to the new unseen data are then determined to be the best match for the new data to be labelled (Nearest Neighbors, 2020). Sklearn with its default value for metric as 'minkowski' and power parameter as '2', is similar to a normal euclidean metric, taking a data point $(x)$ and the predicted point $(y)$ into account with $n$ number of dimensions, Euclidean metric can be denoted by equation-11 (Distance Metric - Sklearn, 2020):

$$f(x, y) = \sqrt{\sum_{i=1}^{n}(x_i - y_i)^2} \qquad (11)$$

If the power parameter is set to '1' with 'minkowski' as default, the algorithm takes Manhattan metric denoted by equation-12 (Distance Metric - Sklearn, 2020):

$$f(x, y) = \sum_{i=1}^{n} \mid x_i - y_i \mid \qquad (12)$$

For arbitrary value of power parameter, the algorithm uses minkowski and it is denoted by equation-13 (Distance Metric - Sklearn, 2020):

$$f(x, y) = (\sum_{i=1}^{n}(\mid x_i - y_i \mid)^q)^{\frac{1}{q}} \qquad (13)$$

Table-XVI shows the metrics output for KNN, with highest F1-score of 0.9984 for CTU-13-scenario 10, along with fit-time as 8345.15 sec, precision as 0.9989, recall as 0.9979 and accuracy as 0.9988.

TABLE XVI
K NEIGHBORS

| Dataset | | Fit Time (s) | Precision | Recall | F1 score | Accuracy |
|---|---|---|---|---|---|---|
| | 1 | 24039.29 | 0.9575 | 0.9391 | 0.9482 | 0.9687 |
| | 2 | 15635.49 | 0.9868 | 0.9728 | 0.9798 | 0.9894 |
| | 3 | 52461.87 | 0.9950 | 0.9985 | 0.9967 | 0.9972 |
| | 4 | 6526.29 | 0.9759 | 0.9350 | 0.9550 | 0.9680 |
| | 5 | 12.51 | 0.9454 | 0.9297 | 0.9375 | 0.9577 |
| | 6 | 2444.85 | 0.9892 | 0.9941 | 0.9917 | 0.9947 |
| CTU-13 | 7 | 7.12 | 0.6799 | 0.5618 | 0.6152 | 0.7081 |
| | 8 | 28655.06 | 0.9633 | 0.9570 | 0.9601 | 0.9864 |
| | 9 | 7335.46 | 0.9457 | 0.9421 | 0.9439 | 0.9582 |
| | 10 | 8345.15 | 0.9989 | 0.9979 | 0.9984 | 0.9988 |
| | 11 | 8.38 | 0.9976 | 0.9982 | 0.9979 | 0.9985 |
| | 12 | 43.16 | 0.9070 | 0.7776 | 0.8373 | 0.8845 |
| | 13 | 8643.61 | 0.9718 | 0.9571 | 0.9644 | 0.9783 |
| ISOT | | 1355.26 | 0.9599 | 0.9763 | 0.9680 | 0.9744 |
| ISCX | | 42.70 | 0.9752 | 0.9808 | 0.9780 | 0.9772 |

Table-XVII shows the hyperparameters to be considered to optimise the performance of K-Neighbors. The default parameters used are [5, 'uniform', 'auto', 30, 2].

TABLE XVII
K-NEIGHBORS - PARAMETER SELECTION

| Parameter | Search Space |
|---|---|
| N neighbors | (low = 1, high = 10) |
| Weights | ['uniform', 'distance'] |
| Algorithm | ['ball tree', 'kd tree'] |
| Leaf size | (low = 1, high = 50) |
| P (power) | (low = 1, high = 5) |

*G. Artificial Neural Networks*

Artificial Neural Networks (ANNs) utilise the concept of biologically-inspired mathematical neurons arranged in groups of units, called layers. Neurons also called perceptrons can be arranged in fully connected layers, where each perceptron has a set of input edges that are connected to a set of output edges produced by the previous layer. Similarly, a perceptron's output edges connect as an input to each perceptron in the next layer. Layers can be constructed in Keras library, using the sequential model to allow various layers (like 'dense' and 'dropout' layers) into a network.

The learning is facilitated by passing input training data to the classifier, whereby each feature vector is assigned an input neuron. Therefore, the input dimension must match feature vector length, which in the case of the extended network flow feature set, is 24. The neural network then passes this data through the layers of fully connected neurons in the network over one or more epochs (iterations), calculating the error of the predictions against a loss metric at each batch, which is then used in the backpropagation to update the weights of the neurons in the network to reduce the cumulative loss.

The feature selection is not performed in NN due to its ability to autonomously discern feature importance during training via backpropagation (Pektas Acarman, 2018). For the ANNs, the 'default hyperparameters' do not exist unlike previously discussed classifiers, hence, two network configurations are utilised with reasonably performing hyperparameters following some initial grid-search exploration with the Keras-Classifier wrappers. Two ANNs are utilised, one with a single hidden layer of 24 units and one with two hidden layers, with 24 units in the first layer and 40 units in the second layer.

*1) Neural Network 1:* Table-XVIII shows the metrics output for ANN#1, with highest F1-score of 0.9176 for CTU-13-scenario-13; along with fit-time as 102.98 sec, precision as 0.9228, recall as 0.9124 and accuracy as 0.9308. When all the scenarios are considered, the performance is poor.

Table-XIX shows the hyperparameters to be considered to optimise the performance of ANN#1. The default parameters used are [5, 100, 100, 'relu', 'sigmoid', 'binary_crossentropy', 'adam'].

*2) Neural Network 2:* Table-XX shows the metrics output for ANN#2, with highest F1-score of 0.7998 for CTU-13-scenario-7; along with fit-time as 27.30 sec, precision as 0.7997, recall as 0.8 and accuracy as 0.9206. When all the scenarios are considered, the performance is poor.

TABLE XVIII
NEURAL NETWORK 1

| Dataset | | Fit Time (s) | Precision | Recall | F1 score | Accuracy |
|---|---|---|---|---|---|---|
| CTU-13 | 1 | 159.02 | 0.6761 | 0.6189 | 0.6462 | 0.6602 |
| | 2 | 105.75 | 0.6902 | 0.6295 | 0.6585 | 0.6649 |
| | 3 | 265.01 | 0.7429 | 0.9038 | 0.8155 | 0.8159 |
| | 4 | 69.69 | 0.6349 | 0.5935 | 0.6135 | 0.6218 |
| | 5 | 14.56 | 0.5341 | 0.5964 | 0.5636 | 0.5757 |
| | 6 | 53.57 | 0.9010 | 0.8580 | 0.8790 | 0.8816 |
| | 7 | 18.66 | 0.6497 | 0.6497 | 0.6497 | 0.7478 |
| | 8 | 175.85 | 0.9125 | 0.8048 | 0.8553 | 0.8787 |
| | 9 | 109.35 | 0.7252 | 0.7401 | 0.7326 | 0.7437 |
| | 10 | 98.38 | 0.6680 | 0.7150 | 0.6907 | 0.6910 |
| | 11 | 19.97 | 0.6463 | 0.6898 | 0.6673 | 0.6677 |
| | 12 | 30.59 | 0.5724 | 0.5779 | 0.5752 | 0.6076 |
| | 13 | 102.98 | 0.9228 | 0.9124 | 0.9176 | 0.9308 |
| ISOT | | 70.52 | 0.4613 | 0.4975 | 0.4787 | 0.4819 |
| ISCX | | 17.17 | 0.6605 | 0.7106 | 0.6846 | 0.6841 |

TABLE XIX
NN1 - PARAMETER SELECTION

| Parameter | Search Space |
|---|---|
| Epochs | (low = 2, high = 10) |
| Batch size | (low = 10, high = 1000) |
| Hidden layers | (low = 10, high = 1000) |
| Layers activation | ['softmax','selu','softplus', 'relu','tanh','sigmoid', 'linear'] |
| Output activation | ['softmax','selu','softplus', 'relu','tanh','sigmoid', 'linear'] |
| Loss function | ['mean_squared_error','mean_absolute_error', 'squared_hinge','categorical_crossentropy', 'binary_crossentropy'] |
| Optimise function | ['SGD','RMSprop','Adagrad','Adadelta','Adam', 'Adamax','Nadam'] |

TABLE XX
NEURAL NETWORK 2

| Dataset | | Fit Time (s) | Precision | Recall | F1 score | Accuracy |
|---|---|---|---|---|---|---|
| CTU-13 | 1 | 598.39 | 0.6715 | 0.5434 | 0.6007 | 0.6133 |
| | 2 | 368.91 | 0.4942 | 0.4999 | 0.4970 | 0.5018 |
| | 3 | 908.09 | 0.4971 | 0.5 | 0.4985 | 0.4988 |
| | 4 | 247.62 | 0.6401 | 0.5951 | 0.6168 | 0.6252 |
| | 5 | 29.89 | 0.6564 | 0.5605 | 0.6047 | 0.6177 |
| | 6 | 141.03 | 0.9354 | 0.6087 | 0.7375 | 0.7397 |
| | 7 | 27.30 | 0.7997 | 0.8 | 0.7998 | 0.9206 |
| | 8 | 527.50 | 0.8579 | 0.6159 | 0.7170 | 0.7366 |
| | 9 | 321.23 | 0.7494 | 0.6193 | 0.6782 | 0.6884 |
| | 10 | 307.27 | 0.6998 | 0.7415 | 0.7201 | 0.7205 |
| | 11 | 31.61 | 0.6551 | 0.6933 | 0.6737 | 0.6741 |
| | 12 | 69.99 | 0.5566 | 0.5535 | 0.5551 | 0.5864 |
| | 13 | 280.27 | 0.5791 | 0.5016 | 0.5376 | 0.5453 |
| ISOT | | 212.00 | 0.4613 | 0.4991 | 0.4795 | 0.4827 |
| ISCX | | 73.01 | 0.6394 | 0.6642 | 0.6516 | 0.6511 |

Table-XXI shows the hyperparameters to be considered to optimise the performance of ANN#2. The default parameters used are [5, 100, 100, 'relu', 0.25, 500, 18, 'relu', 0.25, 'sigmoid', 'binary_crossentropy', 'adam'].

## IX. OPTIMIZATION

### A. Random Search

In a random search methodology, stochastic values are chosen for each hyperparameter type from a set for a given classifier. This method is extremely inefficient and may not result in an optimal hyperparameter set in a reasonable amount

TABLE XXI
NN2 - PARAMETER SELECTION

| Parameter | Search Space |
|---|---|
| Epochs | (low = 2, high = 10) |
| Batch size | (low = 10, high = 1000) |
| Hidden layers | (low = 10, high = 1000) |
| Layers activation | ['softmax','selu','softplus','relu','tanh','sigmoid', 'linear'] |
| Dropout 1 | (low = 0.1, high = 0.5) |
| Hidden layers 2 | (low = 10, high = 1000) |
| Input dim 2 | (low = 2, high = 24) |
| Layers activation 2 | ['softmax','selu','softplus','relu','tanh','sigmoid', 'linear'] |
| Dropout 2 | (low = 0.1, high = 0.5) |
| Output activation | ['softmax','selu','softplus','relu','tanh','sigmoid', 'linear'] |
| Loss function | ['mean_squared_error','mean_absolute_error', 'squared_hinge','categorical_crossentropy', 'binary_crossentropy'] |
| Optimise function | ['SGD','RMSprop','Adagrad','Adadelta','Adam', 'Adamax','Nadam'] |

of computational time for guaranteed improvements (Scikit-learn, 2019a).

### B. Grid Search

In this method of hyperparamets exploration, possible hyperparameter values are arranged in a grid, whereby all possible combinations of hyperparameter sets are constructed and then exhaustively traversed to evaluate each set. Although being slightly better in terms of computational efficiency than the random search, certain instances give low F1 score and take more time (Scikit-learn, 2019a). However, the grid search can explore a more effective combination of hyperparameters outlying from the default values. However, this method is limited to the grid initially defined. Moreover, the resolution of the hyperparameter values is limited to initial values, resulting in a hyperparameter set that may not be near-optimal due to the resolution limitations. Additionally, the efficiency of the grid search method exponentially decreases with the increase in the number of hyperparameters. Large data sets take more time which deteriorates the grid search time efficiency. Table-XXII shows the grid search parameter combinations for all the classifiers.

TABLE XXII
CLASSIFIER GRID SEARCH COMBINATIONS AND TIMES

| Classifier | Grid Search Combinations | Largest Dataset Fit Time (with default hyperparameters) (s) | Total Grid Search Time (hours) |
|---|---|---|---|
| Decision Tree | 1408 | 55.6 | 21.75 |
| Random Forest | 3456 | 80.1 | 76.90 |
| Ada Boost | 840 | 781.9 | 182.44 |
| Linear SVC | 200 | 1514.8 | 84.16 |
| N-Neighbors | 2200 | 52461.9 | 32060.05 |
| Neural Network #1 | 3025260 | 265 | 222692.75 |
| Neural Network #2 | 88942644000 | 908.1 | 22435781949 |

It is worth noticing that the fitting was a single iteration performed (against the largest dataset which is 1 GB in size) with the default hyperparameters for the particular classifier. As changing hyperparameters fundamentally alters the algorithmic operations which can lead to varied fitting times. For example, the RF classifier utilises an n_estimators integer value hyperparameter, which is used to control the number of separate trees utilised in the creation of the forest. As a result, increasing the forest size will also increase processing time required for the classifier to fit the data.

### C. Genetic Algorithm

This research proposes utilising a more efficient method of hyperparameter optimisation. The natural biological evolution tends to favour the fitter individuals in a population, in that they survive longer and breed wherein the genes in the offspring are crossed over from the parents allowing mutation in the genetic makeup. This natural selection process allows the desirable fittest characteristics to be propagated through the future population generations. In the case of hyperparameter optimisation, this research has applied Genetic Algorithm (GA) as a method for converging towards more optimal hyperparameter sets to increase the training efficiency (Lin et al., 2014).

The GA algorithm has proven to be effective in the research done in intrusion detection and malware detection. The authors in the paper (Fatima et al., 2019), experimented to detect malware on an Android platform using GA. The algorithm provides promising results in reducing feature space. The paper also states the capability of GA to work with reduced feature dimensional along with reduction in training time. The paper (Moubayed et al., 2020) worked with optimised RF with GA and found it to be more effective compared to a standalone classifier for botnet detection based on DNS queries, using TI-2016 DNS dataset. In the paper (Resende Drummond, 2018) on adaptive anomaly-based intrusion detection system using GA and profiling, the authors were able to achieve a high detection rate and low false positive rate, when used with CICIDS-2017 dataset. These work and others justifies the use GA in our research, as a potential optimisation approach.

*1) Gene Pools:* Before a GA can operate, a search space is defined for hyperparameter maximum and minimum numerical values along with categorical strings. Therefore, as each hyperparameter takes the form of a 'gene' in the GA, this search space can be thought of as the 'gene pool', defining the states or range of states in which they can exist. This gene pool is not defined as a grid and thus is not constrained to the resolution and intervals present in a grid, in terms of integer and float data-typed hyperparameters.

Although more hyperparameters are available for each of the classifiers in Scikit-Learn library, suitable consideration was taken to ensure that a range of relevant hyperparameters were chosen and that the constraint values of each (depending on data type) are such that the resultant pool is able to encompass a highly dimensional gene pool search space. The gene pool for the classifiers is shown in Table-XXIII.

*2) Operation:* Each individual 'chromosome' in a population consists of a hyperparameter set, with each gene containing the relevant values in the range defined for the hyperparameter in the gene pool. As a result, each chromosome exhibits different lengths and hyperparameter data types (Lin et al., 2014).

In this implementation, a heuristic approach is taken to the initial population generation. This is where the initial population is injected with a known-performance chromosome of default hyperparameter values. This is advantageous to a random initialisation as it gives a place to start if the rest of the population is not initially fit.

In each generation, each member of the population undergoes an evaluation via a fitness function. In this case, the stratified 10-fold cross-validation methodology is utilised to maintain classifier F1 score performance evaluation consistent with the previous experiment. Therefore, the maximisation of this F1 score is desired, and thus in the selection process, the two fittest are chosen as the parents for the next generation. The parents' genes are then crossed over using one central point so that the offspring have the genes of each parent. The population is then filled to the population limit with the offspring and other randomly injected chromosomes. The mutation takes place in a random gene of each chromosome in the population to alter its value, helping to avoid arriving at a local optimum and increase the chances of convergence towards fitter chromosomes.

Figure-4 shows the operation of GA through a flow chart. Algorithm 1 presents a pythonic-style pseudo-representation of the GA class implementation.

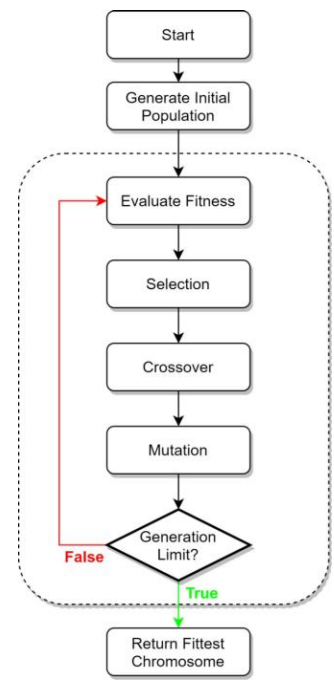

Fig. 4. The operation of Genetic Algorithm

## X. PRODUCT DESIGN

### A. Software Method

Before the design of the product begins, the development model must be chosen. The development model must be suitable for the project that it is being used in the size and resources available. As a result, the V-Model (figure-5) is chosen, as it allows clear and specific validation stages within the testing that can be directly cross-referenced and linked back to the relevant part in the design phase.

TABLE XXIII
THE GENE POOLS DEFINED FOR EACH CLASSIFIER

| Classifier | Gene (Parameter) | Search Space |
|---|---|---|
| Decision Tree | Criterion | ['gini', 'entropy'] |
| | Splitter | ['best', 'random'] |
| | Min_samples_split | (low = 2, high = 5) |
| | Min_samples_leaf | (low = 2, high = 4) |
| | Min_weight_fraction_leaf | (low = 0.0, high = 0.1) |
| | Class_weight | ['balanced', 'None'] |
| Random Forest | N_estimators | (low = 10, high = 200) |
| | Criterion | ['gini', 'entropy'] |
| | Min_samples_split | (low = 2, high = 5) |
| | Min_samples_leaf | (low = 2, high = 4) |
| | Min_weight_fraction_leaf | (low = 0.0, high = 0.1) |
| | Class_weight | ['balanced', 'None'] |
| Ada Boost | N_estimators | (low = 5, high = 100) |
| | Learning_rate | (low = 0.1, high = 1.0) |
| | Algorithm | ['SAMME', 'SAMME.R'] |
| | Random state | (low = 1, high = 50) |
| Linear SVC | Loss | ['hinge', 'squared_hinge'] |
| | Tol (Tolerance) | (low = 1e-5, high = 0.1) |
| | C (Regularisation Parameter) | (low = 1, high = 5) |
| K-Neighbors | N_neighbors | (low = 1, high = 10) |
| | Weights | ['uniform', 'distance'] |
| | Algorithm | ['bal_tree', 'kd_tree'] |
| | Leaf_size | (low = 1, high = 50) |
| | P (Power) | (low = 1, high = 5) |
| Neural Network 1 (24-x1) | Epochs | (low = 2, high = 10) |
| | Batch_size | (low = 10, high = 1000) |
| | Hidden_layers | (low = 2, high = 1000) |
| | Layers_activation | ['softmax', 'selu', 'softplus', 'relu', tanh' 'sigmoid', 'linear'] |
| | Output_activation | ['softmax', 'selu', 'softplus', 'relu', tanh' 'sigmoid', 'linear'] |
| | Loss_function | ['mean_squared_error', 'mean_absolute_error', 'squared_hinge', 'categorical_crossentropy', 'binary_crossentropy'] |
| | Optimiser_function | ['SGD', 'RMSprop', 'Adagrad', 'Adadelta', 'Adam', 'Adamax', 'Nadam'] |
| Neural Network 2 (24-x1-d1-x2-d2-1) | Epochs | (low = 2, high = 10) |
| | Batch_size | (low = 10, high = 1000) |
| | Hidden_layers | (low = 2, high = 1000) |
| | Layers_activation | ['softmax', 'selu', 'softplus', 'relu', tanh' 'sigmoid', 'linear'] |
| | Dropout_1 | (low = 0.1, high = 0.5) |
| | Hidden_layers_2 | (low = 10, high = 1000) |
| | Input_dim_2 | (low = 2, high = 24) |
| | Layers_activation_2 | ['softmax', 'selu', 'softplus', 'relu', tanh' 'sigmoid', 'linear'] |
| | Dropout_2 | (low = 0.1, high = 0.5) |
| | Output_activation | ['softmax', 'selu', 'softplus', 'relu', tanh' 'sigmoid', 'linear'] |
| | Loss_function | ['mean_squared_error', 'mean_absolute_error', 'squared_hinge', 'categorical_crossentropy', 'binary_crossentropy'] |
| | Optimiser_function | ['SGD', 'RMSprop', 'Adagrad', 'Adadelta', 'Adam', 'Adamax', 'Nadam'] |

This method was implemented to allow an explicit structure to be followed, ensuring that all steps are carried out thoroughly in the design stage. The testing can be conducted with relevant areas to relate the tests if any changes are needed, and the process reverts to the traceable design stage that the testing is being conducted at. This allows a dynamic development methodology, only where required.

*B. Design Process*

The design of the IDS engine, according to the V-Model, requires that sufficient functional and non-functional requirements are first defined in the concept phase. To this end, requirements are researched and the inspiration has been taken from other detection systems, such as the operation modes and architecture of the Snort IDS. The design for the IDS that was developed is shown in the figure-6.

The class association diagram is shown in figure-7, that shows the relationship between different classes. The system

---

**Algorithm 1:** Genetic Algorithm for hyper-parameter tuning

---

population = initialise population(populationSize);
chromosomeLength = len(population[0]);
fittestOfGen = [];
**while** *generation < generationLimit* **do**
    fitnessValues = [];
    //Evaluate the fitness F1 scores
    **for** *chromosome in population* **do**
        metrics = CrossValidate(classifier, dataset,
          cv=k);
        f1Score = calcF1Score(metrics);
        fitnessValues.append(f1Score);
    **end**
    //Selection of the fittest chromosomes
    sorted = sort(fitnessValues);
    parent1 = sorted[0];
    parent2 = sorted[1];
    fittestOfGen.append(parent1);
    //Crossover of genes from the fittest parent
    offspring1 = [];
    offspring2 = [];
    **for** *gene in chromosomeLength* **do**
        **if** *gene < (chromosomeLength/2)* **then**
            offspring1.append(parent1[gene];
            offspring2.append(parent2[gene];
        **else**
            offspring1.append(parent2[gene];
            offspring2.append(parent1[gene];
        **end**
    **end**
    //Generate a new population from the offspring
    population = [];
    **for** *index in populationSize* **do**
        **if** *index < populationSize/4* **then**
            chromosome = offspring1.copy();
        **else if** *index < populationSize/2* **then**
            chromosome - offspring2.copy();
        **else**
            chromosome =
            generateRandomChromosome();
        **end**
        population.append(chromosome);
    **end**
    //Mutate a random gene in each chromosome
    **for** *each chromosome in population* **do**
        mutate(chromosome);
    **end**
**end**
bestChromosome = sort(fittestOfGen);
**return** *bestChromosome[0]*;

---

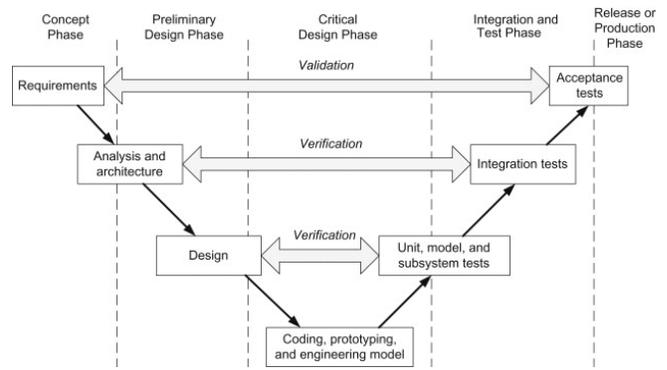

Fig. 5. The V-Model (Fowler, 2015)

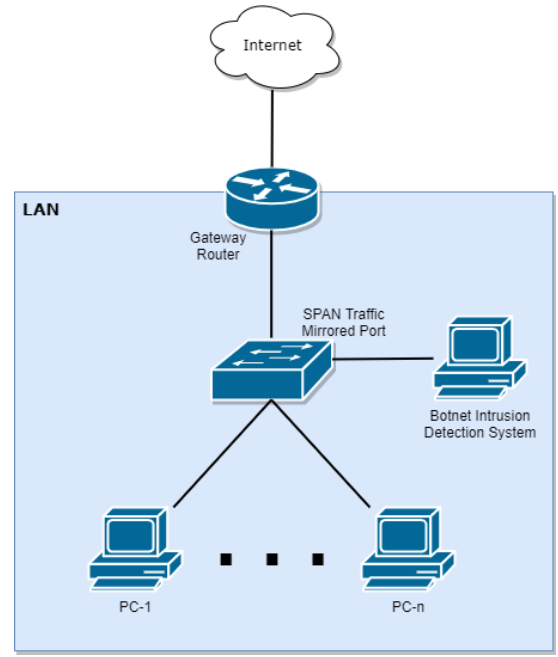

Fig. 6. The Intrusion Detection System Topology

interaction diagram is expressed in figure-8, showing how each object interacts with other objects during operation of the product.

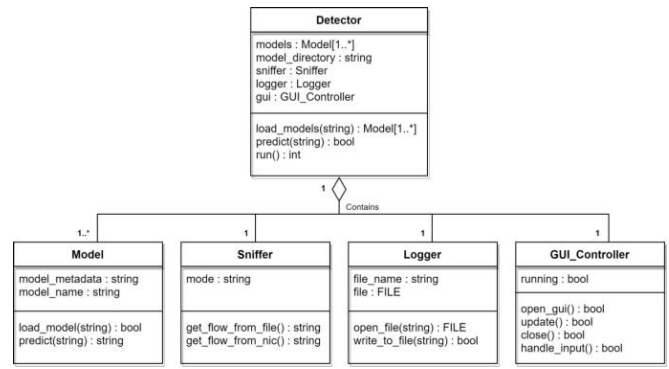

Fig. 7. Class association diagram

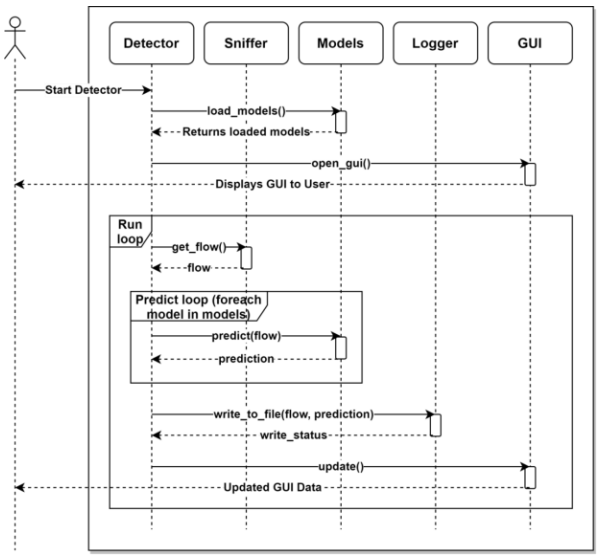

Fig. 8. System Interaction diagram

The high-level architectural design showing the expected input, process and output of data is shown in the figure-9.

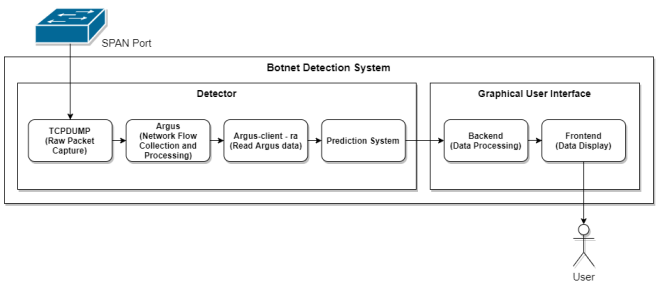

Fig. 9. The Botnet Detection System's high-level architectural design

The GUI intends to be a minimal-interaction data display dashboard, taking predicted data and any alerts from the detector to display in various formats for the user. The detector can operate without the GUI but outputs data to the command line interface.

## XI. PRODUCT IMPLEMENTATION

### A. Network Flow Sniffer

Functional Requirement 1 (F-R-1) as in Table-XXIV mandates that the detection engine should be able to detect and process raw network flow data from the machine's network interface card, generally intended to be connected to a switch through port mirroring or SPAN (Switch Port Analyzer) port.

*1) Operation:* This was originally implemented via a bash script with tcpdump, Argus and the 'ra' client running, with the stdout of each being piped into the stdin of the next program. The final section of the pipe was the stdin of the main detector Python process. This method worked, via a function to read stdin within the sniffer module.

However, the sniffer class should be able to run these processes by itself, without having to rely on the input pipeline being correct in bash script. Therefore, the Python

'subprocess' module was implemented to increase control and integrate the data sniffing and processing components required for successful injection into the sniffer module.

TABLE XXIV
PRODUCT REQUIREMENTS (FUNCTIONAL) SPECIFICATION

| ID | Type | Priority | Requirement |
|---|---|---|---|
| F-R-1 | Functional | Required | The detection engine must be able to process raw network data, received from the machine's NIC from a SPAN port on a switch, into network flow data with the required features present |
| F-R-2 | Functional | Required | The detection engine must load (de-serialise) the serialised machine learning model objects from the files |
| F-R-3 | Functional | Required | The detection engine must be able to predict the classification of each network flow using all of the loaded models |
| F-R-4 | Functional | Required | The detection engine must generate an alert and information for any network flows that have been labelled as 'botnet' by one or more of the models that gave the predictions |
| F-R-5 | Functional | Required | The detection engine must log generated alerts to file |
| F-R-6 | Functional | Required | The detection engine must log labelled network flow data to file |
| F-R-7 | Functional | Required | The detection engine must have the ability to send the labelled flow data to the Graphical User Interface (GUI) |
| F-R-8 | Functional | Required | The GUI must be able to receive data from the detection engine |
| F-R-9 | Functional | Required | The GUI must be able to display alerts |
| F-R-10 | Functional | Required | The GUI must be able to display network flow data |
| F-D-1 | Functional | Desired | The detection engine may have the ability to run standalone (i.e. without the GUI) |
| F-D-2 | Functional | Desired | The detection engine may have the ability to read and process inputted packet capture '.pcap' files (instead of reading from the network) |
| F-D-3 | Functional | Desired | The detection engine may have the ability to read and process inputted network flow files (such as the '.binetflow' or '.csv' dataset files) |
| F-D-4 | Functional | Desired | The detection engine may have the ability to add more bot models to the system (similar to how new rules can be added to the Snort Intrusion Detection System) and dynamically load any new models at load time |
| F-D-5 | Functional | Desired | The GUI may have the ability to display the detection results as a real time network flow activity chart |

*2) Adaptation to Support Multiple Formats of Data Input:* To allow support for multiple formats of data input, i.e. in the case of .pcap or pre-processed network flow files, to achieve requirements F-D-2 and F-D-3 as in Table-XXIV, respectively, the sniffer was designed to utilise an argument, which is initially set to false for normal network sniffing operation, whereby .pcap or network flow files can be specified in the detector via a command-line argument (-r).

If a file is specified, the sniffer is configured at load time as to the mode it is required to run in. This means altering the subprocess commands that are running depending upon the mode.

*3) Subprocess Management:* Some issues were encountered during the development and initial testing whereby the sub-processes would become 'orphans' i.e. they would hang, and keep running perpetually in the background even when the main detector process was terminated.

This was discovered to be a result of having to include the *shell=true* argument in the *Popen* function, as this is required to enable the successful calling and subsequent execution of commands, similar to how they are executed and behave on a command-line shell. This works via the forking of a child process and first executing an sh shell process, which then subsequently calls the command provided as an argument to *Popen*.

The issue subsequently arises during the exiting of the main detection engine python process, whereby the sniffer object's destructor _del_ is called. This destructor originally contained the code below, calling terminate on the subprocess objects, resulting in each of the subprocess' process ID to be sent a SIGTERM signal.

*def del_(self):*
*# The Sniffer object's destructor*
*# Kills the subprocesses via their parent shell PIDs*
*if self.read_from_file == False): if self.tcpdump is not None:*
*# Kill tcpdump*
*os.killpg(self.tcpdump.pid, signal.SIGTERM)*
*if self.argus is not None:*
*# Kill argus*
*os.killpg(self.argus.pid, signal.SIGTERM)*
*if self.ra is not None:*
*# Kill ra*
*os.killpg(self.ra.pid, signal.SIGTERM)*
*print("Deleting sniffer object and any relevant running subprocesses.")*

However, this PID belongs to 'sh' process that is initially forked, which gets properly terminated as expected, but the child process that it forks does not, resulting in the orphaned processes. This is an undesirable characteristic for a final product and caused memory issues during testing, with many orphaned processes maintaining persistence in the background.

To solve this, the orphaned subprocesses must also somehow be sent a SIGTERM signal, and the solution must be properly implemented as to not involve a 'hacky' fix, which could be calling *os.kill(subprocess.pid + 1)*. The subprocess module's *Popen* function was found to support a *preexec_fn* argument, whereby we can specify a system call in the child 'sh' process, after the fork from the parent, but before 'sh' is executed.

This can be used to call *os.setsid*, which creates a new session and process group that results in the calling process to become the process group leader. In this case, the calling process is the sh shell which forks a child process of one of the components, such as the *tcpdump* or *argus* subprocesses, which subsequently inherit the process group ID (PGID) of the calling parent 'sh' process.

As a result, the sh calling parent process and the callee share the same PGID, and thus, all processes within the group can be killed using *os.pgkill*, passing in the PID (as the calling shell process is the process group leader, PID == PGID) of the parent subprocess and SIGTERM. This functionality is subsequently implemented in the Sniffer's _del_ object destructor.

*B. Scalable Model Sets*

Scalable model sets are implemented to meet F-D-5 as in Table-XXIV. This allows new models to be added to the directory and their metadata to be entered into the metadata JSON storage file and they will subsequently be dynamically loaded at runtime – and this is similar functionality to the scalable ruleset design in Snort.

The functionality is implemented in the detector, whereby it finds the models listed in the metadata file and then attempts to find the same name model file in the models directory. Any valid models found are then de-serialised into separate model objects, allowing the detector to maintain an array of successful model objects for its prediction operations.

*C. Graphical User Interface (GUI)*

Electron was chosen over methods such as Tkinter for the GUI development due to its rising popularity for enabling the creation of dynamic desktop-based web-style applications, with a NodeJs server backend and a JavaScript-powered frontend. This allowed a dynamic rendering of CSS-styled HTML pages and interaction with the backend via an exposed send and receive to the frontend sandboxed JavaScript renderer process.

This method enables the power and familiarity of web development to allow desktop applications to be more usable and functional without the overhead of another GUI library. Additionally, this method gives the ability to allow the GUI to be hosted and on another machine on the network for monitoring purposes, as it is disconnected from the running detection engine process, as it runs over a client-server web-socket interaction.

To allow only one page being in memory and being rendered at a time, an API call is invoked at each page load within the renderer process, which requests all the data stored in a data handling class created in the backend to track all received flows and alerts. When this request is received in the main backend process, the event handler calls Data_Controller.get_all_data() which returns all the data stored in the storage arrays and subsequently sends the data back to the frontend. As these are in the backend, they are isolated from the frontend and are not lost when new pages are loaded.

The renderer code for each page then handles the array of received data, as opposed to a single item as was the case in the beginning, and thus iterate through each index and render the specific data correctly. This helps to maintain a perceived persistence and continuity of data between the specific pages available.

*D. WebSockets*

Since the user interface is running as a separate process from the detection engine, the network flow data and alerts must somehow be sent from the detector to the GUI. Initially, Inter Process Communications (IPC's) were considered, but the power, ease and flexibility of a websocket implementation over localhost allowed a quick and reliable method of transferring data using the WebSocket client and server libraries.

The GUI's backend, therefore, hosts a websocket server to listen for incoming connections, and then includes an event handler to handle each message received. Additionally, this method allows for future flexibility, as the GUI could be running on another machine on the network and so the detector on the network edge can connect to the GUI running elsewhere in the network, via its IP address. This is a common use case for dashboard-style monitoring systems, wherein the dashboard is connected over the network to the detector for remote monitoring.

### E. Product (Botnet Detection Software)

The main features of the software are as follows. The network flow classification predictions are done via serialised bio-optimised RF classifier models. The model is scalable and the sniffer supports multiple input types like raw live network sniffing, raw packet capture files (.pcap) and pre-processed (extended feature set) network flow text files (.binetflow). The extensible desktop-based dashboard GUI is made using the Electron framework, with file logging for alerts, labelled network flows and desktop notification alerts.

## XII. DETECTION TESTING

### A. Unit Testing

As per the V-Model development process, first the unit testing was carried out via the use of the unit test module in Python.

A mixture of black and white box testing is utilised. The black box testing intends to ensure that the functions do not fail at a high level, whereas the white box testing ensures that as many statement execution branches in each function are executed and tested for their impacts, whether being assertions of return statement values or against expected variable values after the execution of a specific function.

To help ensure that the black and white box unit tests achieve a large coverage, a report was generated after the script runs. This was to ensure and validate that the unit tests included relevant actions which enable the majority of code statements in all branches to be executed, where reasonably was feasible. Statements that are not directly triggered during the creation of white-box tests generally required excessive modification of the running environment, such as removing other files or processes running on the system.

### B. Integration Testing

The integration testing intends to test high-level functionality, with direct relation to the analysis and architecture design phase, wherein the high-level architecture is defined. These tests can test what white box unit testing do not. The tests were conducted using test data utilising the main run.sh script to enable testing of the various modes available, ensuring that the expected behaviours are present in the actual functionality of both the detector and the GUI.

### C. Acceptance Testing

Acceptance testing directly maps onto the original requirements, and therefore the requirements are the test cases. These acceptance tests are generally what helps to ensure that integration was successful in meeting the required and possibly the desired requirements. In this case, all required and desired specifications were met, with some areas for improvement identified even in this stage for future versions of the software.

## XIII. RESULTS

Table-XXV shows the summary of the results for all classifiers without optimisation using 10-fold cross-validation. From the results, it can be observed that the highest averaged F1 score and accuracy across all of the datasets is achieved by the RF classifier, with an F1-score of 0.9742 and accuracy of 0.9964, closely followed by the DT classifier with an F1-score of 0.9618 and accuracy of 0.9837. The highest average fitting times are present in the kNN classifier which also presents the fourth-highest F1 score.

TABLE XXV
CLASSIFIER PERFORMANCE WITHOUT OPTIMISATION - AVERAGED
ACROSS ALL DATASETS

| Classifier | Fit Time(s) | Precision | Recall | F1 score | Accuracy |
|---|---|---|---|---|---|
| Gaussian Naive Bayes | 3.4279 | 0.5176 | 0.6780 | 0.5781 | 0.5912 |
| Decision Tree | 11.8037 | 0.9597 | 0.9638 | 0.9618 | 0.98372 |
| Random Forest | 23.8155 | 0.9820 | 0.9670 | 0.9742 | 0.9964 |
| Ada Boost | 237.3703 | 0.9734 | 0.9320 | 0.9508 | 0.9725 |
| Linear SVC | 594.6946 | 0.6213 | 0.7003 | 0.6556 | 0.6705 |
| K Neighbors | 10370.4261 | 0.9499 | 0.9279 | 0.9381 | 0.9595 |
| Neural Network 1 (24-x-1) | 87.4093 | 0.6932 | 0.6999 | 0.6952 | 0.7110 |
| Neural Network 2(24-x1-d1-x2-d2-1) | 276.2794 | 0.6595 | 0.5997 | 0.6245 | 0.6387 |

The poor performance exhibited by the Linear SVM classifier may be due to the non-linearly disparate distribution of the botnet samples in the datasets, making it more difficult for the classifier to gain a higher precision.

In their default configuration, the two neural network models do not perform well, both performing under an F1-score of 0.70, which is surprising but can be due to the nature of the networks used. Optimisation of the various hyperparameters will be explored in the next section.

An experiment was carried out using the Snort IDS due to its longstanding presence and usage in the world of network intrusion detection. The default rule sets that are to be utilised are therefore deemed to be more thorough, mature and well developed as a result. The default rule sets that are to be used are known as the 'Community Snort Rules' and are free to use, and open source, under the GNU general public licence version 2 (Snort, 2020). The results are shown in table-XXVI.

It can be observed that there is a range of FPR values for each dataset i.e., from 0.11% to 3.03%. Additionally, the higher FPRs are present in the scenarios that include P2P usage (CTU-Normal-7 and CTU-Normal-12), despite these datasets containing a fraction of the number of packets in the pure web browsing traffic datasets (CTU-Normal-22 and CTU-Normal-31). Therefore, we can tentatively conclude that as more complex scenarios are included in the normal traffic, the FPR rate on even known legitimate traffic will also increase, especially when used in a larger network with more active clients.

TABLE XXVI
SNORT IDS FPR EXPERIMENT RESULTS

| Dataset | Legitimate Activity Type | True Negatives | False Positives (Alerts) | FPR (%) |
|---------|--------------------------|----------------|--------------------------|---------|
| CTU-Normal-7 | P2P D/Ls; Web Browsing | 471239 | 14284 | 3.031158287 |
| CTU-Normal-12 | P2P D/Ls; Web Browsing | 863374 | 14646 | 1.69636797 |
| CTU-Normal-22 | HTTPS Web Browsing | 2621640 | 15928 | 0.607558627 |
| CTU-Normal-31 | HTTPS Web Browsing | 1242661 | 1419 | 0.114190435 |
| - | - | - | Average | 1.36231883 |

The bio-inspired hyperparameter optimisation was executed with a population size of 10 and a generational limit of 10 for the ML classifiers. In the case of two ANNs, a population and generational limit of 15 was chosen due to the much larger gene pools present and thus allowed more hyperparameter set evaluations to take place. Do refer to table-XXVII.

TABLE XXVII
F1 SCORE FOR BIO-OPTIMISED HYPERPARAMETER IMPROVEMENT

| Classifier | Default Hyperparameters | Bio-Optimised Hyperparameters | Improvement |
|-----------|-------------------------|-------------------------------|-------------|
| Decision Tree | 0.961800 | 0.969803 | 0.008003 |
| Random Forest | 0.974219 | 0.977448 | 0.003229 |
| Ada Boost | 0.950865 | 0.958826 | 0.007961 |
| Linear SVC | 0.655630 | 0.740954 | 0.085324 |
| K Neighbors | 0.938194 | 0.940588 | 0.002394 |
| Neural Network 1 (24-x-1) | 0.695254 | 0.837873 | 0.142619 |
| Neural Network 2 (24-x1-d1-x2-d2-1) | 0.624558 | 0.827529 | 0.202971 |

The highest performing classification after hyperparameter optimisation is performed by RF classifier, despite a minor improvement in the F1-score performance after optimisation. In the other classifiers, such as Linear SVM which initially scored very low, the GA optimisation has led to noticeable performance improvement. Additionally, despite the large search spaces present in the gene pools, ANNs resulted in the largest improvements, taking both their F1-score performances above 0.82, with the single hidden layer network having a higher score (0.8379) but lower overall improvement after GA hyperparameter optimisation.

### A. Bio-Optimised Model Validation

When the bio-optimised RF classifier hyperparameter sets are used to train the final models that are to be used in the product for live detection, it is necessary to evaluate their

FPR performance against known legitimate data, as already conducted with Snort in table-XXVI. The FPR is measured using similar datasets as with Snort, and the results are shown in the table-XXVIII.

TABLE XXVIII
THE FALSE POSITIVE RATE COMPARISON OF THE BIO-OPTIMISED RF-GA ON CTU-13 MODELS AGAINST SNORT'S PERFORMANCE

| Dataset | Legitimate Activity Type | FPR of Snort | FPR of optimised RF-GA on CTU-13 Models |
|---------|--------------------------|--------------|------------------------------------------|
| CTU-Normal-7 | P2P Downloads; Web Browsing | 3.031158287 | 0 |
| CTU-Normal-12 | P2P Downloads; Web Browsing | 1.69636797 | 0 |
| CTU-Normal-22 | HTTPS Web Browsing | 0.607558627 | 0 |
| CTU-Normal-31 | HTTPS Web Browsing | 0.114190435 | 0 |
| Average | | 1.36231883 | 0 |

In this case, only the 13 CTU-13 derived RF classifier models are considered, like the models from the ISOT and ISCX datasets can incur larger FPR's due to their more generalised and diverse behaviour datasets. But the clear advantage of our model over Snort IDS performance is clearly seen.

For the highest performing classifier after optimisation, i.e. the RF classifier with GA, the data is presented in table-XXIX. An average accuracy of 99.85% is achieved across all three datasets - CTU-13, ISOT and ISCX, which is a very good achievement. This also suggests that the RF optimisation with GA makes the classifier performance more consistent across different datasets.

Table-XXX, table-XXXI and table-XXXII shows a comparison of our work with similar work. The papers mentioned within the table uses similar datasets. The data for our work shows individual dataset accuracy for CTU-13, ISOT and ISCX datasets. Comparing our work of optimised ML classification (RF-GA) where we had used flow-based data without IP addresses or port numbers, the accuracy and F1-score of our work are better than other work. Our work will also withstand bot evasion attacks like encryption, packet fragmentation etc. as it can only use flow-based data.

The software output when the botnet detection is executed, is shown in figure-10 and figure-11. The figures show the normal traffic flow and bot flow in text and activity graph, with alerts being generated when bot activity is detected.

### XIV. FUTURE WORK

Future work would include the inclusion of more datasets that include more diverse types of bot samples and malicious activities. Moreover, the datasets will be used to label any bot network behaviour from the bot sources that do not specifically define these behaviours individually. Due to this, the models used in the IDS only label detected botnet behaviour and provide an array of the range of possible characteristics that a bot sample exhibits.

The neural network models were found to be not as effective as initially expected, therefore, it is recommended that additional feature extraction techniques be used, or by using

TABLE XXIX

RF Classifier's Bio-optimised Hyperparameters using GA and their performance with 10 Generations and population size of 10

| Dataset | | Pre-Processes size (MB) | Botnet Class (%) | F1-Score (RF) | GA Fitness Chromosome | Precision (RF-GA) | Recall (RF-GA) | F1-Score (RF-GA) | Accuracy (RF-GA) | GA Improvement |
|---|---|---|---|---|---|---|---|---|---|---|
| CTU-13 | 1 | 629 | 1.45 | 0.9758 | [187, 'entropy', 2, 2, 0.0, None] | 0.9825 | 0.9710 | 0.9775 | 0.9987 | 0.001716027 |
| | 2 | 401 | 1.16 | 0.9884 | [100, 'entropy', 2, 2, 0.0, None] | 0.9924 | 0.9869 | 0.9897 | 0.9995 | 0.001315723 |
| | 3 | 1002 | 0.57 | 0.9991 | [187, 'entropy', 3, 1, 0.0, 'balanced'] | 0.9998 | 0.9997 | 0.9994 | 0.9999 | 0.0002534 |
| | 4 | 251 | 0.23 | 0.9848 | [36, 'entropy', 2, 1, 0.0, None] | 0.9875 | 0.9847 | 0.9863 | 0.9998 | 0.001558873 |
| | 5 | 28.9 | 0.69 | 0.9692 | [192, 'entropy', 2, 1, 0.0, 'balanced'] | 0.9899 | 0.9513 | 0.9781 | 0.9993 | 0.008989795 |
| | 6 | 124 | 0.83 | 0.9964 | [70, 'entropy', 2, 1, 0.0 None] | 0.9969 | 0.9985 | 0.9969 | 0.9998 | 0.000433455 |
| | 7 | 25.3 | 0.06 | 0.8513 | [75, 'gini', 3, 1, 0.0, 'balanced'] | 0.9129 | 0.7998 | 0.8685 | 0.9997 | 0.017174908 |
| | 8 | 657 | 0.21 | 0.9727 | [145, 'gini', 3, 2, 0.0, None] | 0.9737 | 0.9742 | 0.9730 | 0.9997 | 0.000323613 |
| | 9 | 464 | 8.86 | 0.9760 | [136, 'gini', 3, 2, 0.0, Noe] | 0.9739 | 0.9738 | 0.9780 | 0.9928 | 0.002041924 |
| | 10 | 291 | 8.12 | 0.9992 | [198, 'entropy', 4, 2, 0.0, 'balanced'] | 0.9995 | 0.9993 | 0.9993 | 0.9997 | 9.47989E-05 |
| | 11 | 23.4 | 7.61 | 0.9991 | [100, 'gini', 2, 1, 0.0, None] | 0.9998 | 0.9997 | 0.9991 | 0.9997 | 3.30176E-05 |
| | 12 | 72.5 | 0.67 | 0.9361 | [169, 'gini', 2, 1, 0.0, None] | 0.9653 | 0.9120 | 0.9453 | 0.9986 | 0.009255644 |
| | 13 | 428 | 2.08 | 0.9838 | [100, 'entropy', 3, 1, 0.0, None] | 0.9852 | 0.9845 | 0.9845 | 0.9987 | 0.000752687 |
| ISOT | | 352 | 7.73 | 0.9887 | [172, 'gini', 3, 1, 0.0, None] | 0.9936 | 0.9857 | 0.9906 | 0.9973 | 0.001888855 |
| ISCX | | 64.7 | 36.61 | 0.9921 | [114, 'gini', 3, 1, 0.0, 'balance'] | 0.9931 | 0.9946 | 0.9947 | 0.9940 | 0.00261135 |
| Average Value | | 320.92 | 5.13 | 0.9742 | NA | 0.9831 | 0.9677 | 0.9774 | 0.9985 | 0.003229605 |

TABLE XXX

Comparison: CTU-13 Dataset

| Author Name | Dataset | Technique | Performance achieved |
|---|---|---|---|
| Chen et al. | CTU-13 | RF | Detection Rate = 93.6% |
| Khan et al. | CTU-13 + ISOT | DT | Accuracy = 98.7% |
| Sinha et al. | CTU-13 | LSTM | Accuracy = 96.2%, TPR = 94.6%, TNR = 96.3% (only for Scenarios 6, 7, 10, 11 & 12) |
| *Proposed Work | CTU-13 | RF & GA | Accuracy = 99.9%, F1-Score = 97.5% |

TABLE XXXI

Comparison: ISCX Dataset

| Author Name | Dataset | Technique | Performance achieved |
|---|---|---|---|
| Bijalwan et al. | ISCX | KNN & DT | Accuracy = 96.41% |
| Alauthaman et al. | ISCX + ISOT | DT & NN | Accuracy = 99.2% |
| Bijalwan | ISCX | Adaboost & DT | Accuracy = 98.36% F1-Score = 98.54% Precision = 98.85% Recall = 98.23% |
| S-K et al. | ISCX | K-means & RF | Accuracy = 99.97% Detection Rate = 98.94% False Alarm Rate = 3.80% |
| Alauthaman et al. | ISCX + ISOT | Reinforcement learning | Accuracy = 98.3% |
| *Proposed Work | ISCX | RF & GA | Accuracy = 99.4%, F1-Score = 99.5% |

TABLE XXXII

Comparison: ISOT Dataset

| Author Name | Dataset | Technique | Performance achieved |
|---|---|---|---|
| Pektas & Acarman | ISOT | RF | F1-Score = 99.0%, Accuracy = 99.5%, Recall = 99.0%, Precision = 99.0% |
| Alauthaman et al. | ISOT + ISCX | Reinforcement learning | Accuracy = 98.3% |
| Mai & Park | ISOT | K-means | Detection Rate = 97.11% |
| Khan et al. | ISOT + CTU-13 | DT | Accuracy = 98.7% |
| Alauthaman et al. | ISOT + ISCX | DT & NN | Accuracy = 99.2% |
| *Proposed Work | ISOT | RF & GA | Accuracy = 99.7%, F1-Score = 99.1% |

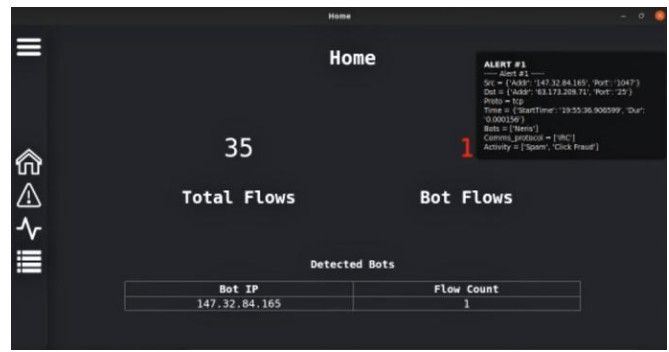

Fig. 10. The software output of botnet detection - Regular and bot traffic

different types of neural networks with data that is processed in a different manner, such as a recurrent neural networks (RNNs) being applied with data that is processed in time windows, to exploit the spatial-temporal nature potentially exhibited in modern botnet network communications.

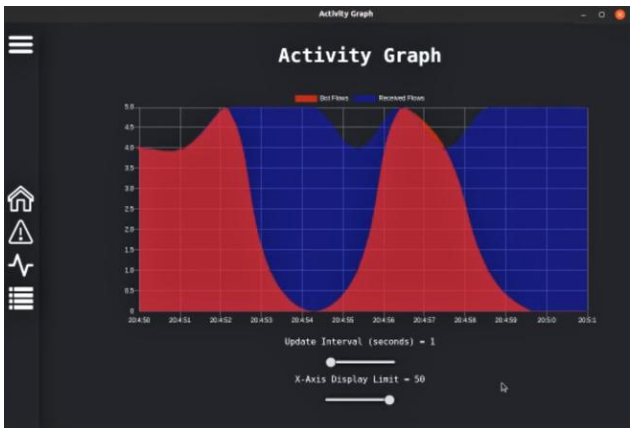

Fig. 11. The software output of botnet detection - Activity graph

Also, an evaluation using a network flow time window can be conducted, with varying the time window length parameter for analysis of the flows. Finally, more bio-optimisation techniques can be explored for the optimisation of the classifier hyperparameters such as firefly, particle swarm optimization, ant colony etc. for comparison purposes. The optimisation times can be compared and it will also be possible to optimise the feature selection.

## XV. CONCLUSION

From the initial analysis of botnets and their operations, it is clear that they pose a large threat to networks globally. They are extremely heterogeneous and possess the ability to autonomously propagate and communicate with a C&C architecture to allow the mass coordination of actions and attacks. Additionally, it is shown that existing IDS are becoming increasingly ineffective to the ever-increasing and effective methods of botnet evasion techniques. Such systems, such as Snort, exhibit large FPRs, even in the presence of known legitimate network traffic. Additionally, the rule sets employed utilise signatures from the communication payload data of bots, which can be obfuscated in various ways to evade this direct payload signature-matching detection mechanism.

The ML methodology successfully enabled network flow features to be extracted and analysed from the publicly available representative, realistic datasets utilised in the evaluation. This methodology proves that botnet C&C communication detection can be facilitated by the characterisation of bot behavioural features using ML classification algorithm, and without the need for signature-based analysis, the network-based behaviour can be analysed. Bio-inspired hyperparameter optimisation was explored using Genetic Algorithm (GA) to allow more efficient convergence to optimal hyperparameters for each classifier. Overall, the RF classifier with GA achieved the highest average F1 score of 97.74% across all of the datasets and accuracy of 99.85% with FPR of 0% against known legitimate datasets when CTU-13 models were trained and serialised with the bio-optimised hyperparameters.

The network flow data is advantageous as it does not incur the privacy concerns that .pcap files present, as they do not include the raw payload data directly. Therefore, they can be publicly distributed more easily and allow further research to be conducted and cross-validated to methodologies in related work. Also, the tools were used to facilitate the feature extraction from raw .pcap files. The ML methodology is then utilised in the product to allow live network flow data feature extraction in the detector's sniffer module.

The product was successfully designed, implemented and tested against all relevant requirements to ensure that the implementation was sufficient. The GUI was developed to present the data to the user in various formats and implements dynamic data loading and pseudo-persistence between pages.

## XVI. ACKNOWLEDGEMENT

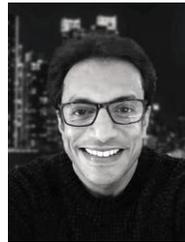

**Biju Issac** received BE degree in Electronics and Communications Engineering, MCA (Master of Computer Applications) degree, and PhD in Networking and Mobile Communications. He is an Associate Professor at Northumbria University, UK, since 2018. He is the Director of Academic Centre of Excellence in Cyber Security Research (ACE-CSR). He has authored more than 100 refereed conference/journal papers, and book chapters. His research interests are in networks, cyber security, AI/machine learning etc.

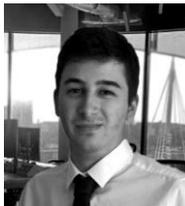

**Kyle Fryer** has completed BSc in Computer Networks and Cyber Security at Northumbria University, UK. Kyle's awards include second prize in the Cyber Security Student Conference (CSSC) and Ede & Ravenscroft Prize for best academic performance at Northumbria University. His research interests are in cybersecurity and machine learning.

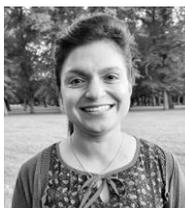

**Seibu Mary Jacob** is an academic staff teaching Mathematics at Teesside University, UK. She took her Bachelor degree in Mathematics (BSc) and Master degree in Mathematics (MSc), along with Post Graduate Diploma in Computer Applications (PGDCA), Bachelor degree in Mathematics Education (BEd) and PhD in Mathematics Education. She has authored more than 20 research publications as book chapters, journal papers and conference papers. She is a member of IEEE, IET, IAENG and IACSIT for many years.